\documentclass[]{agujournal2019}
\pdfoutput=1 
\usepackage{url} %
\usepackage{lineno}
\usepackage[finalnew]{trackchanges} %
\usepackage{soul}

\journalname{}

\begin{document}

\title{Low-level jets and the convergence of Mars data assimilation algorithms}

\authors{Todd A. Mooring\affil{1,*}, Gabrielle E. Davis\affil{1,2}, and Steven J. Greybush\affil{3}}

\affiliation{1}{Department of the Geophysical Sciences, University of Chicago, Chicago, Illinois, USA}
\affiliation{2}{Department of Physics, University of Maryland, Baltimore County, Baltimore, Maryland, USA}
\affiliation{3}{Department of Meteorology and Atmospheric Science, The Pennsylvania State University, University Park, Pennsylvania, USA}
\affiliation{*}{Now at Department of Earth and Planetary Sciences, Harvard University, Cambridge, Massachusetts, USA}

\correspondingauthor{Todd A. Mooring}{tmooring@alum.mit.edu}

\begin{keypoints}
\item Assimilating temperature data \change{into}{in} UK-LMD Mars \change{atmosphere}{climate} model weakens\change{ and}{,} shifts northern \add{winter }low-level jet, but has less effect on GFDL model %
\item Time mean flows generally agree better in the MACDA and EMARS reanalyses than in their associated \change{free-running control simulations}{control runs} %
\item Reanalysis--control run mean state differences suggest that the EMARS control run has smaller biases than the MACDA control run %
\end{keypoints}

\begin{abstract} %
\add{Data assimilation is an increasingly popular technique in Mars atmospheric science, but its effect on the mean states of the underlying atmosphere models has not been thoroughly examined.  The robustness of results to the choice of model and assimilation algorithm also warrants further study.  } %
We investigate \change{the effect of data assimilation on the mean climates of}{these issues using} two Mars general circulation models (MGCMs), with particular emphasis on zonal wind and temperature fields.  %
When temperature retrievals from the Mars Global Surveyor Thermal Emission Spectrometer \add{(TES)} are assimilated into the U.K.-Laboratoire de M\'et\'eorologie Dynamique (UK-LMD) MGCM to create the Mars Analysis Correction Data Assimilation (MACDA) reanalysis, low-level zonal jets in the winter northern hemisphere shift equatorward and weaken relative to a free-running control simulation from the same MGCM.  %
The Ensemble Mars Atmosphere Reanalysis System (EMARS) reanalysis, which \change{assimilates essentially the same}{is also based on TES} temperature retrievals, also shows jet weakening (but \change{not}{less if any} shifting) relative to a control simulation performed with the underlying Geophysical Fluid Dynamics Laboratory (GFDL) MGCM.  %
\remove{For both reanalysis--control simulation pairs, the low-level extratropical jet changes caused by data assimilation are associated with surface pressure changes.  
These surface pressure changes are roughly in geostrophic balance with the wind changes.  }
Examining higher levels of the atmosphere, monthly mean three-dimensional temperature and zonal wind fields are in \change{robustly}{generally} better agreement between the two reanalyses than between the two control simulations.  %
\remove{Furthermore, the MACDA control run typically differs more from MACDA than the EMARS control run differs from EMARS.  }
In conjunction with information about the MGCMs' physical parametrizations, intercomparisons between the various reanalyses and control simulations suggest that \add{overall }the EMARS control run is \change{generally}{plausibly} less biased (relative to the true state of the Martian atmosphere) than the MACDA control run.  %
Implications for future observational studies are discussed.  %
\end{abstract}

\section*{Plain Language Summary}
An increasingly popular way to study Martian weather and climate is to combine atmospheric temperature observations with a computer model (specifically, a general circulation model).  %
The process of combining model and observations is referred to as ``data assimilation'', and the resulting merged data set is referred to as a ``reanalysis''.  %
Several Mars reanalyses have been produced.  %
One advantage of reanalyses is that they include meteorological variables (such as wind) that are not directly observed---however, it is not clear how accurately data assimilation algorithms compute these unobserved variables.  %
Our study investigates this issue using two Mars reanalyses and two model simulations that do not assimilate temperature data.  %
We focus on slowly-varying atmospheric phenomena (timescales from 10 Mars days to a season).  %
Assimilating temperature data into two different general circulation models changes the strength and/or spatial pattern of east-west winds at low altitudes.  %
Furthermore, monthly mean three-dimensional temperature and east-west wind fields agree better between reanalyses than between non-assimilating model simulations.  %
This suggests that the data assimilation process is basically successful.  %
One non-assimilating model simulation has less realistic representations of atmospheric physical processes than the other---we argue that this \change{likely}{plausibly} gives it larger biases relative to the true state of the atmosphere.  %

\section{Introduction}
Data assimilation for the Martian atmosphere has been a subject of research for more than two decades \cite{lewis95,lewis96,houben99} and recent years have seen a proliferation of reanalysis data sets \cite<e.g.,>[]{montabone14,steele14,navarro17,holmes18,greybush19gdj,holmes19,holmes20}.  %
The Martian data assimilation problem must be solved with fewer and different observations than its terrestrial counterpart: to date, \change{the only data directly used to update the atmospheric dynamical state during the assimilation process are infrared temperature retrievals or their underlying radiances}{Mars reanalysis efforts have been highly dependent on infrared temperature retrievals (or at least their underlying radiances) in ways that Earth reanalyses are not} \cite<e.g.,>[]{lee11,montabone14,greybush19gdj}, \cite<cf.>[]{gelaro17,hersbach20}.  %
\change{Other}{This is because other} dynamical information, such as surface pressure or wind observations, is available with only very limited spatial coverage \cite{hinson08,martinez17}. %

From a dynamical perspective, atmospheric temperature structure is most clearly informative about wind fields via thermal wind or similar balance arguments \cite<e.g.,>[]{banfield04}.  %
However, thermal wind is at best a theory of the vertical wind shear---it cannot constrain the absolute wind at the surface and is also expected to break down in the tropics.  %
Thus although the large-scale near-surface and tropical atmospheric circulations are basic features of the Martian climate system, it is not obvious how well they are estimated by data assimilation systems \cite{lewis96,lewis97,hoffman10}.  %
Nor are the simulations of these features by free-running Mars general circulation models (MGCMs) easy to validate.  %

Here we \change{investigate}{begin to address these product quality issues by investigating} how assimilating temperature retrievals into MGCMs changes their climatological mean states, with particular emphasis on zonal winds.  %
To explore the robustness of our results, we examine two different reanalyses and their associated control simulations---the control simulations differ from the reanalyses primarily by not assimilating temperature retrievals.  %
The use of two reanalysis--control run pairs also allows us to expand on previous investigations \cite{waugh16,greybush19te} of whether different data assimilation systems are able to converge on a single atmospheric state.  %
Ultimately we are able to draw some \add{tentative }conclusions about the \remove{likely }quality of the reanalyses and control simulations, even without using any independent validation data.  %

\change{T}{The main body of t}his paper is divided into four major sections.  %
We summarize the reanalysis data sets and control simulations in section~\ref{sec:datasets}.  %
Results on the low-level zonal mean jets are presented in section~\ref{sec:lowlev_jets}, while the vertical and meridional structure of the zonal mean temperature and zonal wind fields is examined in section~\ref{sec:zonal_means}.  %
The extent to which data assimilation converges the time mean states of the two MGCMs is addressed more formally in section~\ref{sec:meanflow_convergence}.  %
A summary and discussion of implications for future observational work concludes the paper\add{, and three appendices present results of sensitivity tests and additional statistical details}.  %

\section{Reanalysis and control simulation data sets}
\label{sec:datasets}
We use the Mars Analysis Correction Data Assimilation version 1.0 \cite<MACDA,>[]{montabone14} and Ensemble Mars Atmosphere Reanalysis System version 1.0 \cite<EMARS,>[]{greybush19gdj} reanalyses, both of which assimilate \change{essentially the same temperature data---retrievals}{temperature retrievals} from the Mars Global Surveyor Thermal Emission Spectrometer \cite<TES,>[]{conrath00}.
This gives the two reanalyses similar temporal extents: MY24 $L_s$ 141$^\circ$ (103$^\circ$) to MY27 $L_s$ 86$^\circ$ (102$^\circ$) for MACDA (EMARS), where the Mars years (MY) and seasonal dates are defined using the \citeA{clancy00} calendar.  However, occasional gaps in the availability of TES retrievals mean that the reanalyses are not constrained by observations throughout the full lengths of these periods.  Ten intervals in which the reanalyses are thought to be poorly constrained are excluded from our study, generally following Table~S1 of \citeA{mooring15}.  (Two more such intervals occur near the beginning of the EMARS data set, but are rendered irrelevant by our choice to ignore the period prior to MY24 $L_s$ 135$^\circ$. \add{We also do not use the MY28-33 segment of EMARS based on Mars Climate Sounder retrievals}.)

The two reanalyses are underpinned by substantially different MGCMs and data assimilation algorithms.  MACDA is based on the U.K.-Laboratoire de M\'et\'eorologie Dynamique (UK-LMD) MGCM with a spectral dynamical core \cite{forget99}.  The MACDA version of this model was integrated with a horizontal resolution of T31 and 25 sigma levels \cite{montabone06}, and the MACDA output data are available on a 5$^\circ$ latitude-longitude grid.  EMARS uses a version of the Geophysical Fluid Dynamics Laboratory (GFDL) MGCM with a finite-volume dynamical core on a latitude-longitude grid \cite<e.g.,>[]{hoffman10}.  The horizontal resolution of this model is $\sim$5$^\circ$ latitude $\times$ 6$^\circ$ longitude, and it has 28 hybrid sigma-pressure levels.

MACDA assimilates temperature retrievals using the analysis correction method \cite{lewis07}, which updates the model state every dynamical timestep (480 times per sol---a sol is a Martian mean solar day, $\sim$1.03 Earth days).  %
In contrast, EMARS assimilates temperature retrievals 24 times per sol using an ensemble Kalman filter \cite{hoffman10,zhao15}.  %
The MACDA data set is available 12 times per sol \cite{montabone14}, while EMARS analyses are available 24 times per sol \cite{greybush19gdj}.  \change{The (Mars) hourly background forecasts, which the ensemble Kalman filter updates to form the analyses, are also available 24 times per sol.  Because the EMARS background forecasts share the same horizontal grid as the output files from the EMARS control simulation, while the EMARS analyses do not, we opted to use the background forecasts (rather than the analyses per se) as EMARS's observationally-constrained representation of the Martian atmospheric state.}{Note that the publicly available EMARS output consists of both analyses and short (1 Mars hour) background forecasts---although many atmospheric variables are available as forecasts only, the pressure, temperature, and wind variables needed for this study are available as both analyses and forecasts and we opt to use the former as they are (slightly) more observationally constrained.}

The free-running control simulations are essentially identical to their associated reanalyses, \change{with two major exceptions}{except that by definition they do not assimilate temperature retrievals}.  \change{First, by definition the control simulations do not assimilate temperature retrievals.  Second, although the MACDA control simulation covers the same multi-Mars year period as MACDA, the EMARS control simulation has a length of only $\sim$1.2 Mars years (MY24 $L_s$ 103$^\circ$ to MY25 $L_s$ 180$^\circ$)---in contrast to the $\sim$3 Mars years of EMARS itself.}{It is important to emphasize that the EMARS control simulation used in this study (version 1.02) is substantially longer than the (version 1.0) control simulation described in} \citeA{greybush19gdj}, \add{which covered only $\sim$1 Mars year of the TES era}.  The MACDA and EMARS control simulations will hereinafter be referred to as MCTRL and ECTRL, respectively.

Even though the control simulations are not constrained by temperature retrievals, they can still be identified with specific Mars years and seasons because their dust fields are time-dependent and constrained by observations.  %
For MACDA and MCTRL, TES-based column opacities are assimilated using the analysis correction method \cite{montabone14}---however, this particular version of the UK-LMD MGCM does not transport dust so the ``forecast model'' underlying the dust opacity assimilation is simply persistence.  %
Given the analyzed column opacities, MACDA and MCTRL distribute the opacity in the vertical using a Conrath-like distribution \cite{conrath75,montabone06}.
In contrast, the three-dimensional dust fields in EMARS and ECTRL evolve under the influences of wind advection and sedimentation \cite{greybush19gdj}.  Agreement with observational data is maintained by nudging the column opacities towards the time-dependent %
dust maps of \citeA{montabone15}\add{, which can also be considered a simple form of data assimilation}.  %
Note that the \citeA{montabone15} dust maps for the period in question are based on retrievals not only from TES, but also from the Thermal Emission Imaging System (THEMIS) on Mars Odyssey.

\section{Low-level zonal jets}
\label{sec:lowlev_jets}
\begin{figure}[b]
\vspace{-.7in}
\hspace{-.875in}
\includegraphics[scale=1]{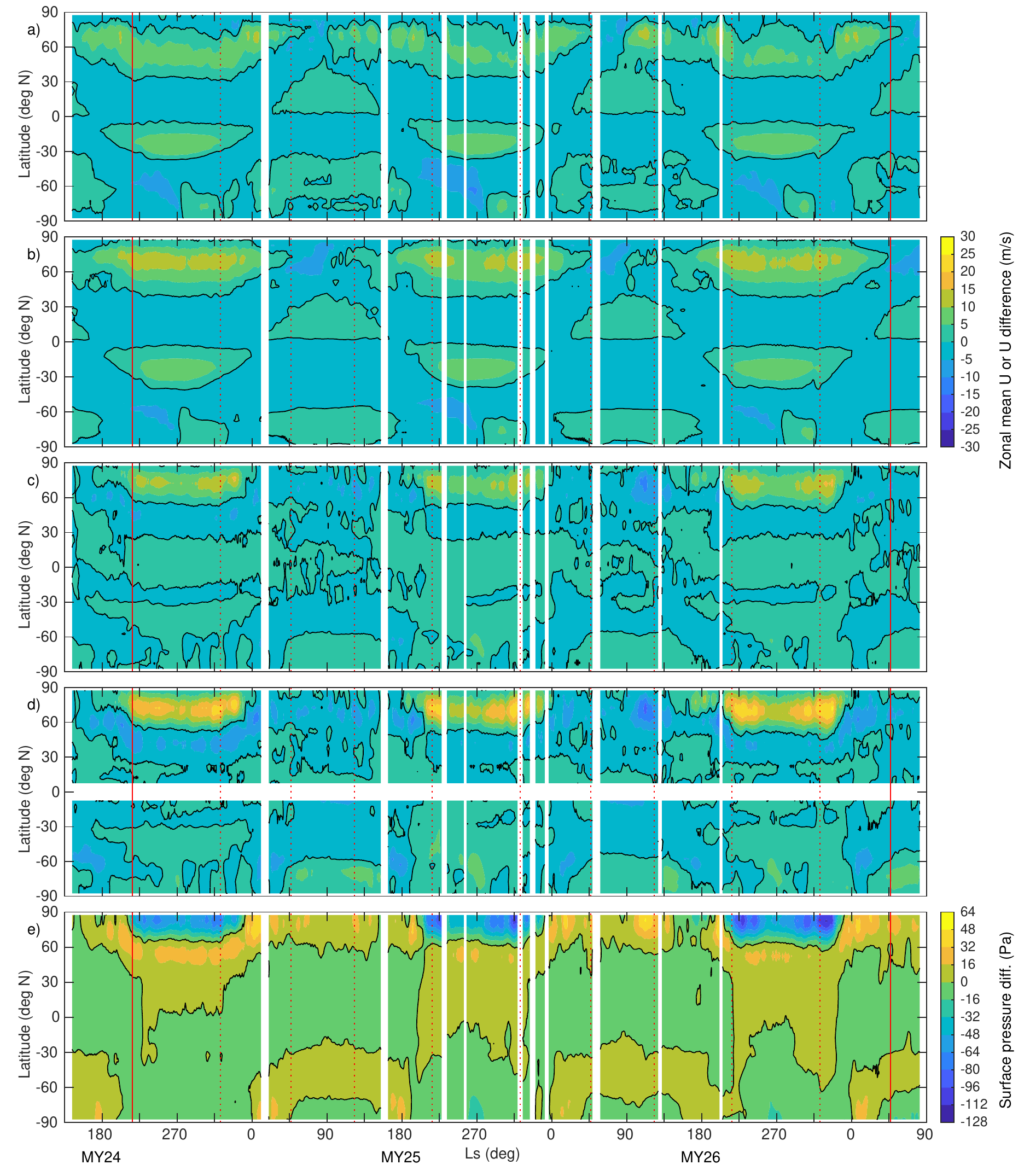}
\vspace{-.4in}
\caption{\linespread{0.8} \selectfont Zonal mean zonal winds on the $\sigma$ = 0.991 level ($\sim$90 m above ground) for MACDA (a) and MCTRL (b).  Differences between MCTRL and MACDA are shown in (c-e)---$\sigma$~=~0.991 zonal winds in (c), surface geostrophic zonal winds in (d) and surface pressures in (e).  The time range shown is MY24 $L_s$ 135$^\circ$ to MY27 $L_s$ 90$^\circ$.  All fields have been smoothed with a 10-sol running mean. The surface geostrophic wind difference in (d) was computed from surface pressure and temperature data from the lowest model level, following equation~4 of \citeA{mooring15}.  \add{(Geostrophic wind differences are not plotted within 7.5$^\circ$ of the equator.)}  The global mean atmospheric mass difference at each timestep was removed before plotting (e).  The black line is the zero contour, notable gaps in the availability of TES retrievals are masked out in white, and the limits of the seasons used in Figures~\ref{fig:basic_states} and~\ref{fig:basic_states_summer_spring} are marked with red lines.\remove{  See main text for more details.}} 
\label{fig:macda_lowlev}
\end{figure}

We begin our comparison of the reanalysis and control run circulations by examining seasonally-resolved zonal mean zonal winds on the $\sigma=0.991$ ($\sim$90 m above ground) level in MACDA and MCTRL.  %
Northern (southern) winter solstice occurs at $L_s$ 270$^\circ$ (90$^\circ$), and focusing initially on the northern hemisphere during its local winter we see that the peak strength of the extratropical zonal jet is lower in MACDA (Figure~\ref{fig:macda_lowlev}a) than in MCTRL (Figure~\ref{fig:macda_lowlev}b).  %
The control run jet also tends to be farther poleward than its reanalysis counterpart.  %
This point is clarified in Figure~\ref{fig:macda_lowlev}c, which shows the difference between the MCTRL and MACDA fields.  %
Figure~\ref{fig:macda_lowlev}c also reveals qualitatively similar behavior in the southern hemisphere near local winter solstice, which was masked in the previously mentioned figure panels by the usually weaker southern winter extratropical near-surface jet.  %
Generally similar wind results are found on the $\sigma=0.900$ ($\sim$1.1 km above ground) level (\ref{app:altitude}, Figure~\ref{fig:macda_hilev}).  %
Furthermore, the MACDA--MCTRL jet differences are associated with differences in zonal mean surface pressure (Figure~\ref{fig:macda_lowlev}e).  %
The differences in surface pressure shown in Figure~\ref{fig:macda_lowlev}e are qualitatively consistent with geostrophic balance and the wind differences shown in Figure~\ref{fig:macda_lowlev}c, although the surface geostrophic zonal wind differences are \add{often }stronger than the actual wind differences at $\sigma=0.991$ (Figure~\ref{fig:macda_lowlev}d). %

\begin{figure}[b]
\vspace{-.7in}
\hspace{-.875in}
\includegraphics[scale=1]{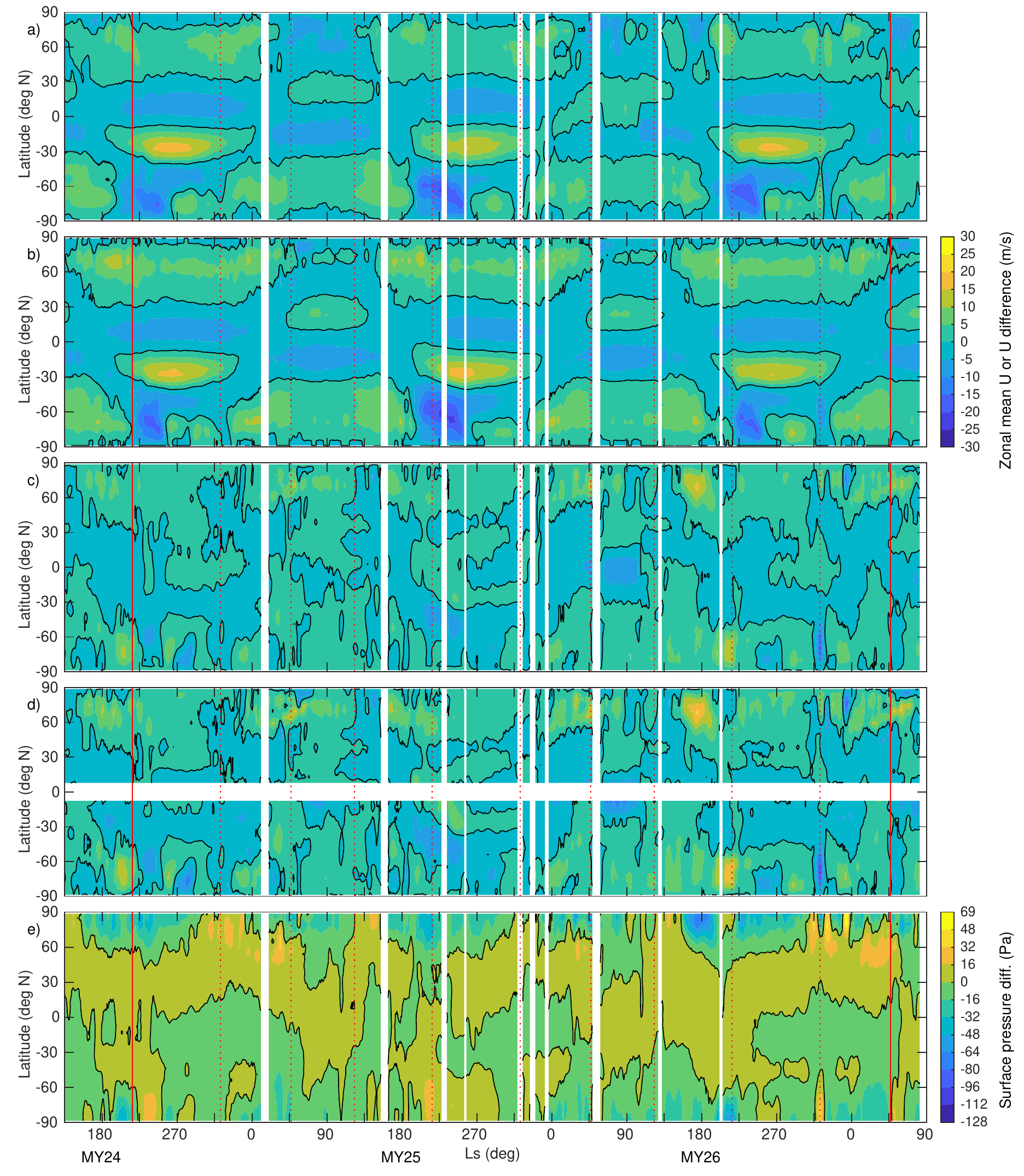}
\vspace{-.3in}
\caption{As Figure~\ref{fig:macda_lowlev} but for EMARS and ECTRL.  Zonal winds in (a-c) are evaluated on the model level with $\sigma$ $\approx$ 0.988 ($\sim$120 m above ground).  \remove{The time range shown is MY24 $L_s$ 135$^\circ$ to MY25 $L_s$ 195$^\circ$.}} %
\label{fig:emars_lowlev}
\end{figure}

A comparable analysis of EMARS and ECTRL yields notably different results (Figure~\ref{fig:emars_lowlev}).  %
\change{Although the relatively short length of ECTRL limits the temporal extent of a direct comparison between the two data sets, there}{There} is a tendency for the assimilation of temperature data to weaken the extratropical winter jets \add{near 60$^\circ$ latitude }in both hemispheres (Figure~\ref{fig:emars_lowlev}a-\change{b}{c}).  %
However, in contrast to the situation with the UK-LMD MGCM, \remove{data assimilation has no obvious effect on the position of the zonal jets---the EMARS--ECTRL jet difference field (Figure~2c) lacks the clear extratropical dipolar structures seen in its MACDA--MCTRL counterpart (Figure~1c).}\add{data assimilation has no obvious effect on the position of the zonal jets---the clear extratropical dipolar structures seen in the MACDA--MCTRL jet difference field }(Figure~\ref{fig:macda_lowlev}c) \add{are absent or greatly weakened in its EMARS--ECTRL counterpart} (Figure~\ref{fig:emars_lowlev}c).  %
The maximum magnitudes of control--reanalysis \add{northern winter} jet differences appear to be smaller for EMARS--ECTRL than for MACDA--MCTRL\remove{, at least in the northern hemisphere} (Figures~\ref{fig:macda_lowlev}c and~\ref{fig:emars_lowlev}c).  %
As with the UK-LMD MGCM, comparable results are found when winds are evaluated on a model level with $\sigma\approx 0.905$ ($\sim$1.0 km above ground, Figure~\ref{fig:emars_hilev}).  %
\change{Note also that}{Interestingly,} the data assimilation effect on surface pressure gradients has a different seasonality in the GFDL MGCM than in the UK-LMD MGCM---for example, the structure of the EMARS--ECTRL northern hemisphere pressure difference field (Figure~\ref{fig:emars_lowlev}e) changes substantially during the \add{MY24 and MY25} $L_s$ 225$^\circ$--315$^\circ$ seasonal interval\add{s} but the corresponding MACDA--MCTRL field does not (Figure~\ref{fig:macda_lowlev}e).  %
\add{Furthermore, even the typical sign of the data assimilation effect on northern hemisphere summer pressure gradients differs between the GFDL and UK-LMD MGCMs (Figure~\protect\ref{fig:macda_lowlev}e and~\protect\ref{fig:emars_lowlev}e).} %
However, as for MACDA--MCTRL the EMARS--ECTRL surface geostrophic wind differences (Figure~\ref{fig:emars_lowlev}d) effectively capture the actual patterns of low-level zonal wind differences.  %

\add{Finally, we note in passing a dubious and previously undocumented feature of EMARS.  Starting near MY26 $L_s$ $\sim$0$^\circ$ and continuing to $L_s$ $\sim$105$^\circ$, the zonal near-surface winds are typically westerly at the equator (}Figures~\ref{fig:emars_lowlev}a and~\ref{fig:emars_hilev}a\add{).  This is in stark contrast to the winds at this season in MY25 and MY27 of EMARS, and in all Mars years of ECTRL (}Figures~\ref{fig:emars_lowlev}b and~\ref{fig:emars_hilev}b\add{).  The abrupt transition to easterly winds near MY26 $L_s$ 105$^\circ$ is coincident with the switch from the second to the third of the separately-initialized EMARS production streams }\cite{greybush19gdj}\add{, and is therefore almost certainly an artifact.  Although the pre-transition westerlies are clearly an outlier relative to the rest of EMARS and all of ECTRL, a more definitive assessment of whether the pre-transition westerlies or post-transition easterlies are more realistic requires additional research.}

\section{Latitude-pressure structure of zonal mean fields}
\label{sec:zonal_means}
Unfortunately, there are very few observations directly sensitive to wind in the lower atmosphere of Mars---anemometers on a handful of landers \cite<e.g.,>[]{martinez17}, geostrophic winds from radio occultations \cite<e.g.,>[]{hinson99}, and arguably cloud-tracked winds from orbiter imagery \cite{wang03c}.  The potential for a direct validation of reanalysis-based winds is thus limited.  However, we can much more readily evaluate the extent to which MACDA and EMARS converge to the same solution---as they should, to the extent that the assimilated data can effectively constrain and correct biases in the MGCM states.  Although our ultimate goal in this paper is to conduct a novel intercomparison of the three-dimensional time mean states of MACDA, EMARS, and their control simulations, we will lead into such an analysis with an examination of zonally-averaged time mean fields.  %

\remove{Because of the strong seasonality of the Martian atmosphere and the limited temporal extent of ECTRL, we will analyze 668 sols of data (almost exactly 1 Mars year) beginning at MY24 $L_s$ 144$^\circ$.  This period is divided into four 167-sol seasons essentially centered on solstices and equinoxes: boreal winter, spring, summer, and autumn are thus $L_s$ 216$^\circ$--322$^\circ$, 322$^\circ$--46$^\circ$, 46$^\circ$--123$^\circ$, and 123$^\circ$--216$^\circ$, respectively.  The 668 sols beginning at MY24 $L_s$ 144$^\circ$ are marked in Figures~1 and~2 with solid red lines, while borders between the seasons are marked with dashed red lines.  (The boreal autumn mean combines data from both MY24 and MY25.)}

\add{Because of the strong seasonality of the Martian atmosphere, for this analysis we will divide the Martian annual cycle into four seasons of nearly equal length and essentially centered on the solstices and equinoxes. More specifically, we define boreal winter, spring, summer, and autumn as $L_s$ 216$^\circ$--322$^\circ$, 322$^\circ$--46.7$^\circ$, 46.7$^\circ$--123$^\circ$, and 123$^\circ$--216$^\circ$. The 2.5 Mars year interval from MY24 $L_s$ 216$^\circ$ to MY27 $L_s$ 46.7$^\circ$ then consists of exactly 10 seasons---three (two) realizations each of boreal winter and spring (summer and autumn). In Figures}~\ref{fig:macda_lowlev} and~\ref{fig:emars_lowlev}\add{, the beginning and end of this 2.5 Mars year period are marked with solid red lines and the borders between individual seasons are marked with dashed red lines.}

\begin{figure}[b]
\vspace{-.3in}
\hspace{-.875in}
\includegraphics[scale=1]{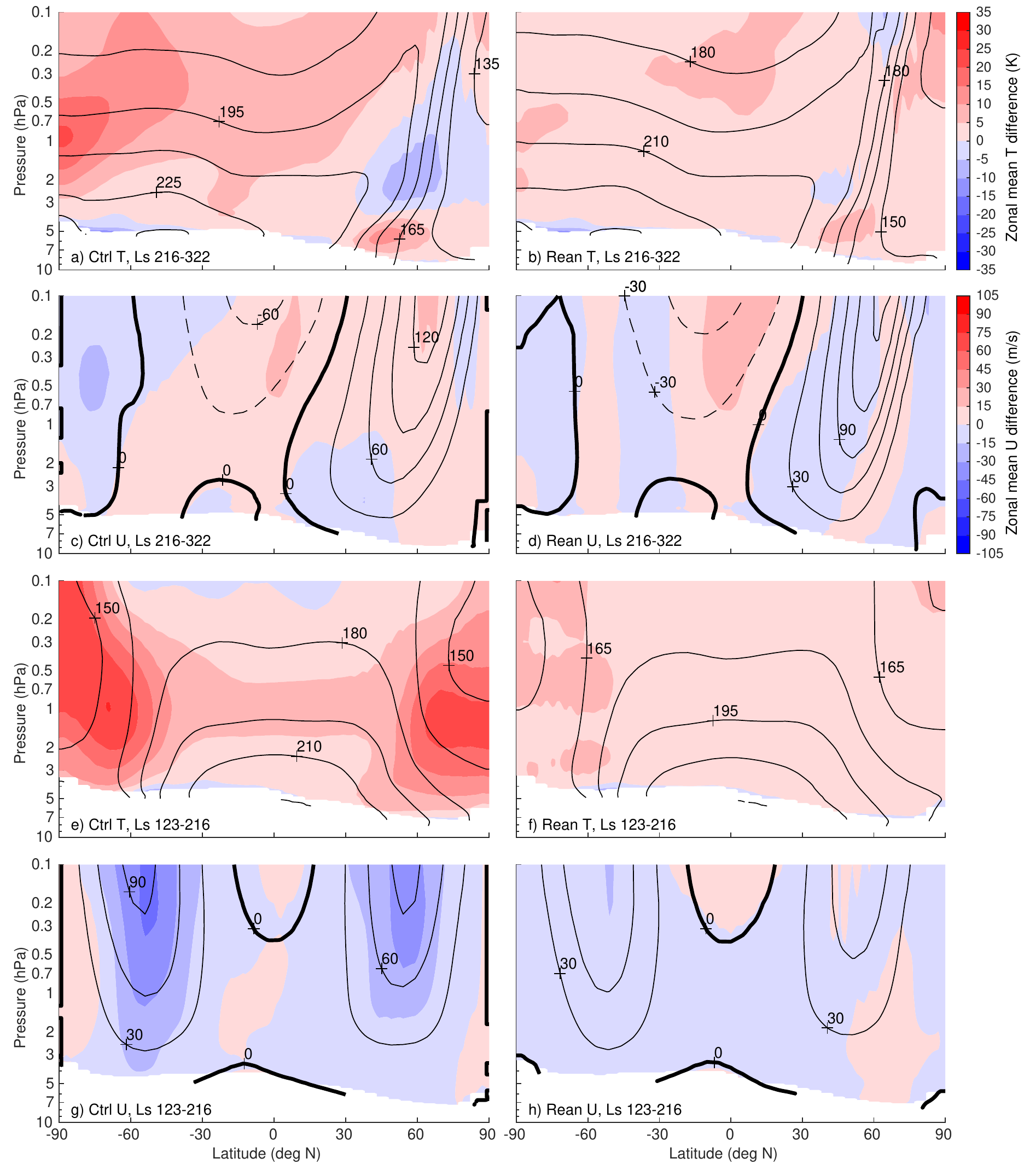}
\vspace{-.3in}
\caption{\linespread{0.8} \selectfont Agreement between reanalysis and free-running control zonal mean temperature and zonal wind fields for boreal winter (a-d) and autumn (e-h).  Black contours in the left (right) column show full fields from ECTRL (EMARS)\add{, with the zero contour marked with a heavy black line}.  Red and blue shading in the left (right) column shows MCTRL minus ECTRL (MACDA minus EMARS).  \change{See main text for more details.}{Interannual means are computed across all available realizations of each season, while each single-Mars year seasonal mean is computed from four monthly means.  The months have lengths of $\sim$41.8 sols, as described in section}~\ref{sec:meanflow_convergence}.}
\label{fig:basic_states}
\end{figure}

\begin{figure}[b]
\vspace{-.3in}
\hspace{-.875in}
\includegraphics[scale=1]{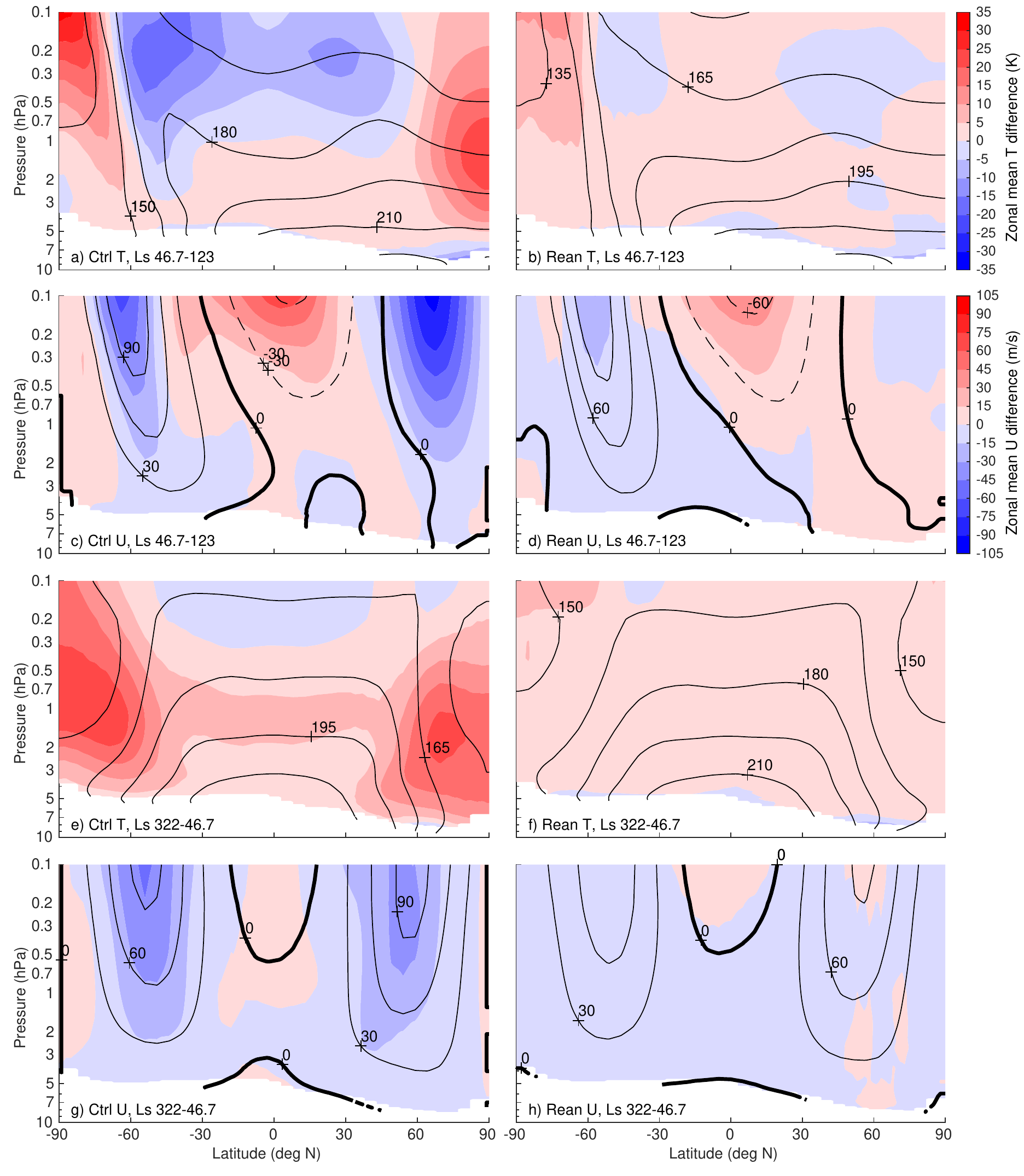}
\vspace{-.3in}
\caption{As Figure~\ref{fig:basic_states}, but for boreal summer (a-d) and spring (e-h).}
\label{fig:basic_states_summer_spring}
\end{figure}

An initial examination of the vertical and meridional structures of zonal mean temperature and zonal wind fields suggests that assimilating TES temperature retrievals brings the UK-LMD and GFDL MGCM states closer together.  %
Results for $L_s$ 123$^\circ$--216$^\circ$ and 216$^\circ$--322$^\circ$ are shown in Figure~\ref{fig:basic_states}. %
Although ECTRL is able to basically reproduce the seasonal variations seen in EMARS (black contours), the disagreements (red and blue shading) between MCTRL and ECTRL (Figure~\ref{fig:basic_states}a, c, e, g) tend to be larger than those between MACDA and EMARS \add{except possibly for the $L_s$ 216$^\circ$--322$^\circ$ zonal winds} (Figure~\ref{fig:basic_states}b, d, f, h).  %
While MACDA is often warmer than EMARS (Figure~\ref{fig:basic_states}b, f), maximum temperature disagreements for these seasons are larger in the free-running control simulations than in the reanalyses: for example, MCTRL can be more than 20 K warmer than ECTRL in the polar regions (Figure~\ref{fig:basic_states}a, e).  %
\change{Due to thermal wind balance, these temperature disagreements imply larger jet disagreements between the control simulations than between the reanalyses}{These patterns of temperature disagreement are associated with jet disagreement due to thermal wind balance---such disagreements are often but not always larger in the control simulations, especially in the extratropics for $L_s$ 123$^\circ$--216$^\circ$ and in high southern latitudes for $L_s$ 216$^\circ$--322$^\circ$} (Figure~\ref{fig:basic_states}c, d, g, h).  %
A tendency of temperature assimilation to converge the UK-LMD and GFDL MGCM mean states is also seen for the other two seasons (Figure~\ref{fig:basic_states_summer_spring}).  %
\change{While}{Although} the patterns of difference between MCTRL and ECTRL are much alike in the two equinox seasons (Figures~\ref{fig:basic_states}e, g and \ref{fig:basic_states_summer_spring}e, g), they appear to disagree more strongly during boreal summer than during boreal winter (Figures~\ref{fig:basic_states}a, c and \ref{fig:basic_states_summer_spring}a, c). %

\section{Convergence of three-dimensional mean fields}
\label{sec:meanflow_convergence}
We can obtain more systematic and quantitative results by computing root mean square (RMS) differences between the various free-running MGCM and reanalysis data sets.  For some three-dimensional time mean field $F$, let us denote the \add{(area- and mass-weighted, assuming hydrostatic balance)} RMS difference between data sets $X$ and $Y$ as $rmsd(X,Y)$.  \change{We}{More precisely, we} define $rmsd(X,Y)$ by
\begin{linenomath*}
\begin{equation}
rmsd(X,Y)=\sqrt{\frac{\int_{\phi_R}\int_0^{2\pi}\int_{p_t}^{p_b}\left(F_X-F_Y\right)^2dp\left(\cos\phi\,d\lambda\right)\,d\phi}{\int_{\phi_R}\int_0^{2\pi}\int_{p_t}^{p_b}dp\left(\cos\phi\,d\lambda\right)\,d\phi}}
\label{eq:rmsd}
\end{equation}
\end{linenomath*}
where $F_X$ and $F_Y$ are field $F$ from data sets $X$ and $Y$, $p_t$ and $p_b$ are the pressures of the top and bottom of the region of interest, and $\phi_R$ denotes the latitude range(s) of interest---the domain over which the meridional integral is taken need not be continuous.  %

It is worth explaining our definition of the time mean.  \add{Our interest is in the mean state of the atmosphere, so the averaging period must be chosen long enough to average out the transient eddies.  However, an excessively long averaging period would needlessly erase information about any shorter-term changes in the mean state.}  \change{For these RMS difference calculations, we opt to split the $\sim$1 Mars year period we are analyzing into 16 months, by subdividing each of the four seasons into four months of 41.75 sols apiece.}{We will again analyze the 2.5 Mars year interval from MY24 $L_s$ 216$^\circ$ to MY27 $L_s$ 46.7$^\circ$ and will attempt to balance these two competing goals by dividing each of the 10 seasons defined in section~}\ref{sec:zonal_means}\add{ into four months with approximately equal lengths of $\sim$41.8 sols.  We then take time means over each of the 40 such months---although because we exclude periods not well constrained by TES data (section~}\ref{sec:datasets}, Figures~\ref{fig:macda_lowlev} and~\ref{fig:emars_lowlev}\add{), four of these monthly means are based on less than 30 sols of data apiece.}  %
Time averaging over \change{periods this long}{$\sim$41.8-sol months} should suffice to suppress most transient eddy variability \cite<e.g.,>[]{banfield04,mooring15}---to the extent that this goal is achieved, any improvement in the agreement of monthly means due to assimilation of TES temperature retrievals should come from correcting the MGCMs' time mean biases and not from synchronizing their unforced variability.  Indeed, repeating the analyses \change{using 12 months of 55$\frac{2}{3}$ sols}{with a month redefined as one-third of a season ($\sim$55.7 sols)} did not qualitatively change the main results (\ref{app:12months}).

We evaluate equation~\ref{eq:rmsd} for each of the \change{16}{40} months for two choices of $F$, 10 (overlapping) spatial regions of interest, and all six possible unique pairs of data sets.  %
The fields used are temperature and zonal wind, and $p_t$ is either 0.1 or 3 hPa.  %
$p_b$ is a spatially-varying monthly mean surface pressure.  %
Specifically, for each location it is computed as the minimum of the four individual data set (MACDA, MCTRL, EMARS, ECTRL) monthly means after the data sets have all been interpolated to a single grid.  %
The choice of $p_t$~=~0.1 hPa excludes altitudes above those directly influenced by TES temperature profile assimilation \cite{lewis07}, while using $p_t$~=~3 hPa emphasizes the lower part of the atmosphere for greater comparability to the results in section~\ref{sec:lowlev_jets}.  %

The 10 spatial regions are formed by combining the two pressure ranges with five latitude ranges: global (90$^\circ$S--90$^\circ$N), tropics (30$^\circ$S--30$^\circ$N), northern and southern hemisphere extratropics (30$^\circ$--90$^\circ$N and 30$^\circ$--90$^\circ$S, respectively) and all extratropics (the union of northern and southern extratropics).  %
While the various latitude ranges are clearly not all independent, using multiple latitude bands is helpful for checking the robustness of the results and investigating whether the effectiveness of temperature assimilation in converging different MGCM mean states varies meridionally.  %

By comparing the relative sizes of the different $rmsd(X,Y)$ we provide support for two major claims:
\begin{enumerate}
\item Assimilating temperature retrievals into the MGCMs brings their monthly mean states into better agreement %
\item ECTRL is \change{probably}{plausibly} less biased (with respect to the true monthly mean states of the Martian atmosphere) than MCTRL %
\end{enumerate}
Knowledge of the actual values of the $rmsd(X,Y)$ is not necessary to support these claims---instead, the results are presented in Table~\ref{tab:rmsd_num_months} in terms of the numbers of months (out of \change{16}{40} possible) for which various inequalities involving the six $rmsd(X,Y)$ are satisfied.  %
\add{For compactness of notation, in these inequalities we will denote MACDA, MCTRL, EMARS, and ECTRL as $M_R$, $M_C$, $E_R$, and $E_C$, respectively.} %

We support the first claim by examining the inequality
\begin{linenomath*}
\begin{equation}
rmsd(M_C,E_C)<rmsd(M_R,E_R)
\label{eq:ctrl_rean}
\end{equation}
\end{linenomath*}
Physically, this inequality will be satisfied if the free-running control simulations are in {\it better} agreement than the reanalyses are (for the given month, field, and region of interest).  %
If this is the case, it means that assimilating TES temperature retrievals does {\it not} systematically bring the monthly mean states of the UK-LMD and GFDL MGCMs together---contrary to the impression created by Figures~\ref{fig:basic_states} and~\ref{fig:basic_states_summer_spring}.  %
\begin{table}[b]
\vspace{-.3in}
\caption{Relative sizes of RMS differences between reanalyses and control simulations}
\hspace{-.6in}
\begin{tabular}{|lll|r|r|r|r|}
\hline
Field & Domain & Meridional & $rmsd(M_C,E_C)<$ & $rmsd(M_R,M_C)<$ & $rmsd(E_R,M_C)<$ & $rmsd(M_R,E_C)<$ \\
      & top (hPa) & domain & $rmsd(M_R,E_R)$ & $rmsd(E_R,E_C)$ & $rmsd(E_R,E_C)$ & $rmsd(M_R,M_C)$ \\
\hline
T & 0.1 & Global &   & 7 &   & 14 \\
T & 0.1 & Tropics & 1 & 5 &   & 6 \\
T & 0.1 & SH extratropics &   & 12 & 2 & 11 \\
T & 0.1 & NH extratropics &   & 12 & 7 & 24 \\
T & 0.1 & All extratropics &   & 8 & 3 & 18 \\
\hline
T & 3 & Global &   & 11 & 2 & 7 \\
T & 3 & Tropics & 3 & 4 &   & 3 \\
T & 3 & SH extratropics & 5 & 12 & 3 & 8 \\
T & 3 & NH extratropics &   & 15 & 6 & 15 \\
T & 3 & All extratropics &   & 12 & 3 & 7 \\
\hline
U & 0.1 & Global & 1 & 10 &   & 19 \\
U & 0.1 & Tropics & 17 & 18 &   &   \\
U & 0.1 & SH extratropics & 8 & 6 & 1 & 14 \\
U & 0.1 & NH extratropics & 1 & 9 & 6 & 28 \\
U & 0.1 & All extratropics &   & 4 &   & 28 \\
\hline
U & 3 & Global & 4 & 16 & 2 & 13 \\
U & 3 & Tropics & 18 & 26 & 5 &   \\
U & 3 & SH extratropics & 13 & 11 & 3 & 4 \\
U & 3 & NH extratropics & 8 & 11 & 6 & 20 \\
U & 3 & All extratropics & 2 & 11 & 1 & 15 \\
\hline
\multicolumn{7}{l}{This table contains information about relative levels of agreement between the various reanalysis and}\\
\multicolumn{7}{l}{control simulation data sets.  We denote the RMS difference between data sets $X$ and $Y$ as $rmsd(X,Y)$.}\\
\multicolumn{7}{l}{The left three columns name the variable being analyzed and the region over which RMS differences}\\
\multicolumn{7}{l}{are being computed.  The right four columns contain the results, expressed as the number of months (of \change{16}{40}}\\
\multicolumn{7}{l}{total) for which the inequality given at the top of each column is satisfied.  Zeros have been}\\
\multicolumn{7}{l}{omitted for clarity.  As an example of how to read the table, the large number of \change{(implied) zeros}{values $\ll$40} in}\\
\multicolumn{7}{l}{the $rmsd(E_R,M_C)<rmsd(E_R,E_C)$ column means that EMARS is in robustly better agreement}\\
\multicolumn{7}{l}{with ECTRL than with MCTRL.}\\
\end{tabular}
\label{tab:rmsd_num_months} %
\end{table}

In practice, equation~\ref{eq:ctrl_rean} is generally not satisfied---Table~\ref{tab:rmsd_num_months} indicates that equation~\ref{eq:ctrl_rean} is true in at most \change{6}{18 and often many fewer} of the \change{16}{40 total} months.  %
If consideration is restricted to the global or all-extratropics meridional regions, the inequality is satisfied for at most \change{one}{four} month\add{s}.  %
These results strongly suggest that assimilation of the same temperature retrievals into UK-LMD and GFDL MGCM simulations tends to bring together not merely their instantaneous weather conditions, but also their climates as measured by monthly means---a more formal statistical analysis suggests that if data assimilation had no effect whatsoever on the MGCMs' monthly mean states, it is unlikely that these results would have been obtained (\ref{app:stats}).  %
\add{Perhaps unsurprisingly, the tendency for data assimilation to converge the monthly means appears stronger for temperature than for zonal wind---for a given region, equation~}\ref{eq:ctrl_rean}\add{ is always satisfied in at least as many months for zonal wind as for temperature.}

We begin to support the second claim by examining
\begin{linenomath*}
\begin{equation}
rmsd(M_R,M_C)<rmsd(E_R,E_C)
\label{eq:uk_gfdl}
\end{equation}
\end{linenomath*}
If satisfied, this inequality indicates that the UK-LMD reanalysis--control run pair is in better agreement than the GFDL reanalysis--control run pair.  %
\change{However, a}{A}cross all of the different field--region combinations equation~\ref{eq:uk_gfdl} is satisfied in \change{no more than 8}{as many as 26} months (Table~\ref{tab:rmsd_num_months}).  %
\change{Excluding the $p_t$ = 3 hPa}{However if the} tropical zonal wind\remove{s} case\add{s are excluded}, it is never satisfied in more than \change{5}{16} months.  %
\change{Evidently}{This is evidence (albeit not always very strong) that} EMARS and ECTRL are generally in better agreement than MACDA and MCTRL\add{, at least outside the tropics}.  %
One possible explanation for this \change{finding}{apparent result} is that ECTRL is less biased (relative to the truth) than MCTRL.  %
However, we cannot immediately dismiss the possibility that the ECTRL biases are comparable to or larger than those of MCTRL but that the EMARS ensemble Kalman filter is simply less effective than the MACDA analysis correction scheme at adjusting the mean state of a biased MGCM.  %
We can separate these possibilities using the additional inequalities
\begin{linenomath*}
\begin{equation}
rmsd(E_R,M_C)<rmsd(E_R,E_C)
\label{eq:emars}
\end{equation}
\end{linenomath*}
and
\begin{linenomath*}
\begin{equation}
rmsd(M_R,E_C)<rmsd(M_R,M_C)
\label{eq:macda}
\end{equation}
\end{linenomath*}
The former (latter) characterizes how well the two control simulations verify against EMARS (MACDA).  %
If ECTRL were clearly superior to MCTRL (in the sense of verifying better against both reanalyses) then equation~\ref{eq:macda} would often be satisfied and equation~\ref{eq:emars} would not be.  %
Likewise, if MCTRL were superior equation~\ref{eq:emars} would often be satisfied and equation~\ref{eq:macda} would not be.  %
Alternatively, if both reanalyses were strongly biased toward their underlying MGCMs both equation~\ref{eq:emars} and equation~\ref{eq:macda} would be only rarely satisfied.  %

The results support the idea that ECTRL is generally less biased than MCTRL---equation~\ref{eq:emars} is satisfied in \change{3}{7} months at most but equation~\ref{eq:macda} is satisfied in as many as \change{13}{28} months (Table~\ref{tab:rmsd_num_months}).  %
Furthermore, for \change{a given field and region}{most field--region combinations} equation~\ref{eq:macda} is \remove{always }satisfied \change{for at least as many}{in more} months \change{as}{than} equation~\ref{eq:emars}\add{---the exceptions are the tropical zonal wind cases}.  %
Statistical analysis suggests that these results\change{ }{---at least for the spatial regions that have $p_t$~=~0.1 hPa and are not wholly tropical---}are unlikely to be explicable as pure interval variability\change{---in other words,}{.  In practice, this implies that} ECTRL and MCTRL have distinct climates and are not simply different realizations of internal variability from a single climate (\ref{app:stats}).  %
Note also that for \change{some}{certain} field--region combinations both equation~\ref{eq:emars} and equation~\ref{eq:macda} are rarely or never satisfied\change{---this is}{,} consistent with the idea that the reanalyses have some tendency to inherit the climates of their underlying MGCMs.  %
\add{This phenomenon is particularly prominent in the tropics.} %

Indeed, there are physical reasons to expect ECTRL to be less biased than MCTRL.  %
Although both control simulations have their column dust opacities constrained to follow similar observational data sets, the constraint method used for ECTRL is more \change{fully}{clearly} consistent with the physics of dust transport in the atmosphere as described in section~\ref{sec:datasets}.  %
Previous work suggests that this should yield more realistic temperatures \cite{wilson08}.  %
Also, the Martian atmosphere features water ice clouds which are thought to substantially affect the thermal structure and circulation \cite<e.g.,>[]{wilson08,mulholland15}.  %
Parameterizations of the radiative effects of water ice clouds have been developed for both the GFDL and UK-LMD MGCMs \cite<e.g.,>[]{hinson04,mulholland15}.  %
They are used in the EMARS--ECTRL version of the GFDL model, but not in the MACDA--MCTRL version of the UK-LMD model \cite{forget99,montabone14,greybush19gdj}.  %
Since the physical parameterizations of ECTRL are a priori more realistic than those of MCTRL, it would be unsurprising if the output of the former simulation were closer to the truth.  %

\section{Summary and discussion}
\label{sec:conclusion}
We have presented a systematic intercomparison of slowly-varying components of the circulation in two Mars reanalyses and their associated free-running control simulations.  %
The reanalyses assimilate essentially the same temperature retrievals, but via very different algorithms and into two distinct Mars general circulation models.  %
Nevertheless, the three-dimensional monthly mean temperature and zonal wind fields are generally in better agreement for the reanalyses than for the control simulations.  %
This suggests a certain robustness of Mars reanalyses to the choice of MGCM and assimilation algorithm, in agreement with \citeA{waugh16} and \citeA{greybush19te}.  %

We devote particular attention to the low-level extratropical zonal mean zonal jets.  %
Assimilating temperature retrievals into the UK-LMD MGCM to create MACDA tends to weaken the northern hemisphere winter jet and to shift it equatorward.  %
Roughly similar shift behavior is found for southern hemisphere winter as well.  %
Weakening of low-level winter jets also results when temperatures are assimilated into the GFDL MGCM, although the overall effect is more subtle than for the UK-LMD MGCM.  %
Furthermore, changes in surface pressure gradients occur in response to temperature assimilation---these are qualitatively consistent with geostrophic balance, most evidently for northern hemisphere winter in the UK-LMD MGCM.  %

Finally, we have produced evidence that (at least in \change{a global}{an} average sense) the EMARS control simulation is less biased than the MACDA control simulation.  %
Note that this result is not guaranteed to hold for individual meridional or vertical regions, such as the tropics or pressures $>$3~hPa---indeed, our results are consistent with the idea that the reanalyses inherit biases from their underlying MGCMs for at least some regions and fields.  %

Our results suggest that the low-level zonal jets of MGCMs may be biased and that similar biases might be shared across multiple MGCMs.  %
Studies of low-level circulations in the Martian atmosphere would thus benefit from collection of additional data more sensitive to near-surface wind or pressure fields.  %
Technological options for collecting such data include lander networks \cite<e.g.,>[]{harri17}, radio occultation constellations \cite<e.g.,>[]{kursinski12}, and orbiting wind lidars \cite<e.g.,>[]{cremons20}.  %
Alternatively, it may be possible to derive improved constraints on low-level zonal geostrophic winds from existing radio occultation and/or lander data.  %
Further MGCM experiments and reanalysis diagnostic studies are also needed to understand the origins of the MGCM--reanalysis and inter-reanalysis disagreements documented here.  %

\appendix
\section{Sensitivity of low-level jets to altitude}
\label{app:altitude}

\begin{figure}[b]
\vspace{-.7in}
\hspace{-.875in}
\includegraphics[scale=1]{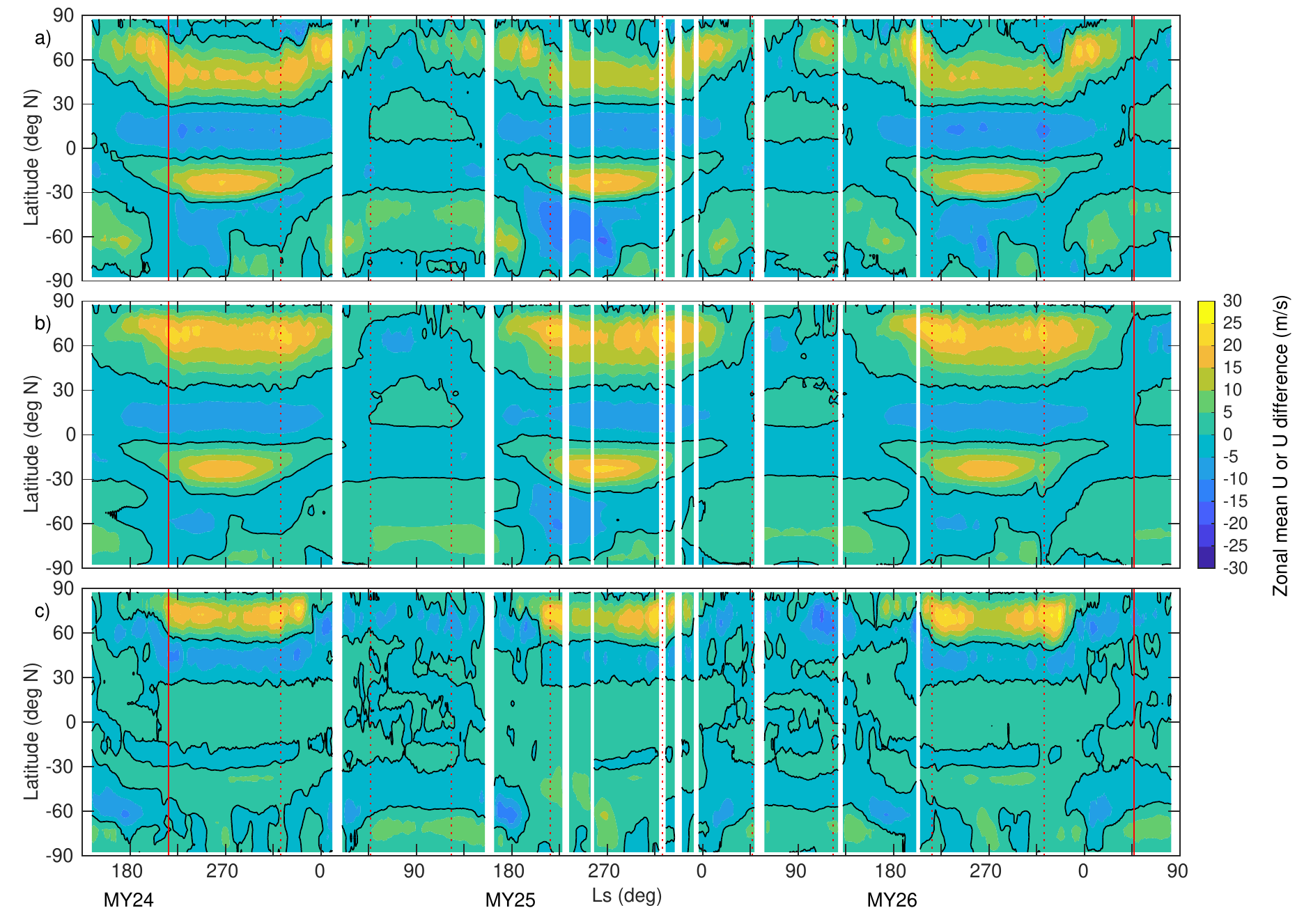}
%
\caption{As Figure~\ref{fig:macda_lowlev}\add{a-c}, but for the $\sigma$ = 0.900 level ($\sim$1.1 km above ground).\remove{  The information in (d-e) is exactly the same as in the corresponding panels of Figure~1, but is reproduced here for clarity.}}
\label{fig:macda_hilev}
\end{figure}

\begin{figure}[b]
\vspace{-.7in}
\hspace{-.875in}
\includegraphics[scale=1]{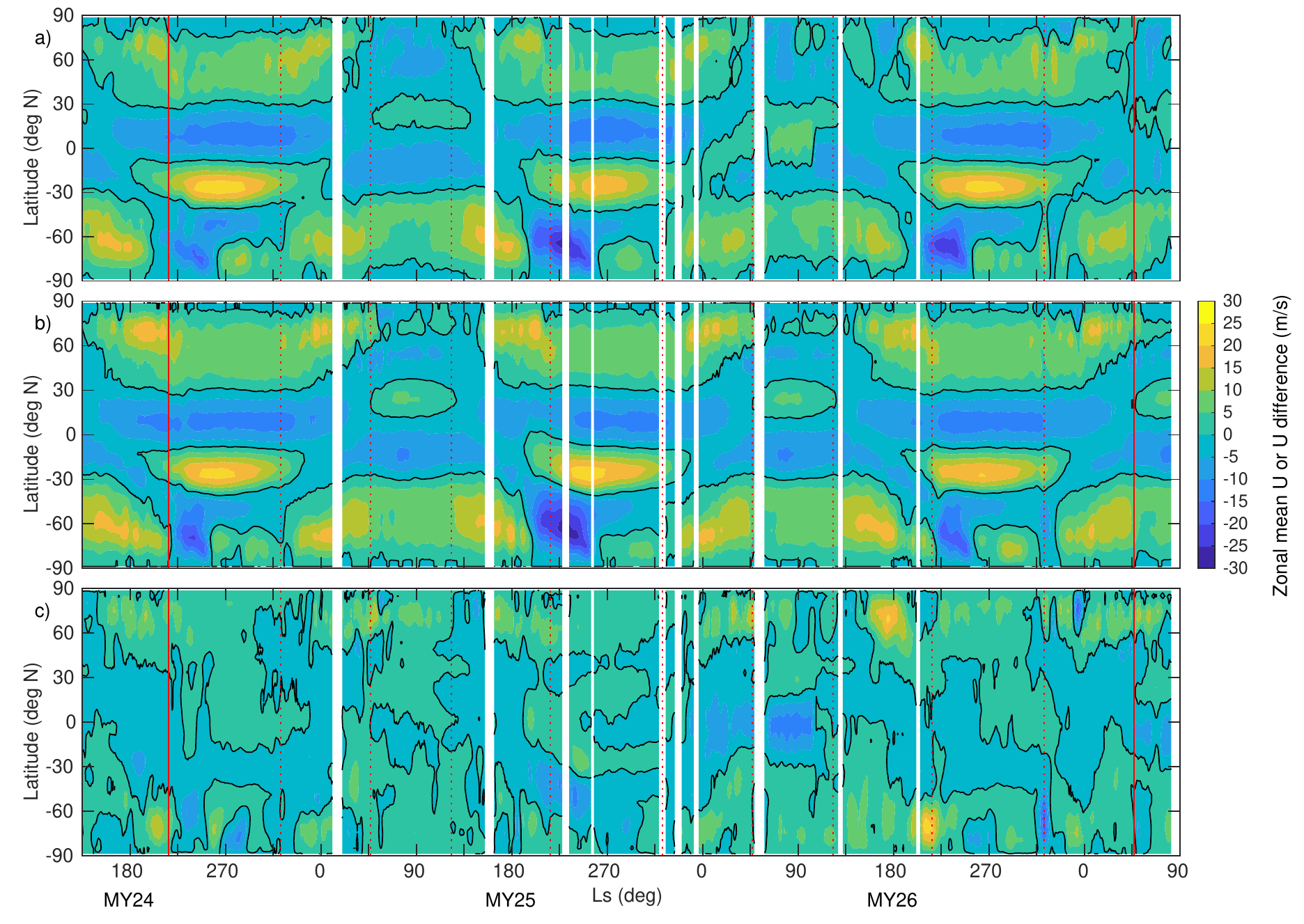}
%
%
\caption{As Figure~\ref{fig:emars_lowlev}\add{a-c}, but on the model level with $\sigma$ $\approx$ 0.905 ($\sim$1.0 km above ground).\remove{  The information in (d-e) is exactly the same as in the corresponding panels of Figure~2, but is reproduced here for clarity.}}
\label{fig:emars_hilev}
\end{figure}

Our primary examination in section~\ref{sec:lowlev_jets} of the seasonal and meridional variations of low-level zonal jets evaluated them on model levels roughly 0.1~km above ground (Figures~\ref{fig:macda_lowlev} and~\ref{fig:emars_lowlev}).  %
To make sure our findings are not strongly sensitive to this arbitrary altitude choice, we repeated the analysis on model levels roughly 1~km above ground and show the results in Figures~\ref{fig:macda_hilev} and~\ref{fig:emars_hilev}.  %
Jet behavior at the two altitudes is basically similar.  %

\section{RMS difference calculation with \change{12}{30} \change{55$\frac{2}{3}$}{$\sim$55.7}-sol months}
\label{app:12months}

To verify that our results concerning the three-dimensional time mean states are robust to the somewhat arbitrary choice of averaging period, we repeated the root mean square (RMS) difference calculations \change{using 12 months of 55$\frac{2}{3}$ sols apiece instead of the standard 16 41.75-sol months}{with each of the 10 seasons divided into three months of $\sim$55.7 sols apiece}.  
\remove{Note that these new longer months evenly divide the 668-sol analysis period and thus do not evenly divide the four seasons defined in the main paper---in other words, some of the new months overlap the season boundaries.}Tables~\ref{tab:rmsd_num_months_12} and~\ref{tab:pvals_12} are the \change{55$\frac{2}{3}$}{$\sim$55.7}-sol month counterparts of Tables~\ref{tab:rmsd_num_months} and~\ref{tab:pvals_16}, respectively.  \add{While the exact quantitative results differ from those obtained with the $\sim$41.8-sol months, the qualitative summary text in section~}\ref{sec:meanflow_convergence}\add{ is based on all four tables and as such is robust to the choice of a $\sim$41.8-sol or $\sim$55.7-sol averaging period.}\remove{We see that the most important results obtained with 41.75-sol months are reproduced here: [table struck]}
\begin{table}[b]
\vspace{-.3in}
\caption{Relative sizes of RMS differences between reanalyses and control simulations}
\hspace{-.6in}
\begin{tabular}{|lll|r|r|r|r|}
\hline
Field & Domain & Meridional & $rmsd(M_C,E_C)<$ & $rmsd(M_R,M_C)<$ & $rmsd(E_R,M_C)<$ & $rmsd(M_R,E_C)<$ \\
      & top (hPa) & domain & $rmsd(M_R,E_R)$ & $rmsd(E_R,E_C)$ & $rmsd(E_R,E_C)$ & $rmsd(M_R,M_C)$ \\
\hline
T & 0.1 & Global &   & 5 &   & 12 \\
T & 0.1 & Tropics & 1 & 3 &   & 5 \\
T & 0.1 & SH extratropics &   & 7 & 1 & 8 \\
T & 0.1 & NH extratropics &   & 11 & 5 & 17 \\
T & 0.1 & All extratropics &   & 8 & 3 & 14 \\
\hline
T & 3 & Global &   & 7 &   & 6 \\
T & 3 & Tropics & 2 & 1 &   & 2 \\
T & 3 & SH extratropics & 4 & 9 & 2 & 7 \\
T & 3 & NH extratropics &   & 10 & 7 & 10 \\
T & 3 & All extratropics &   & 8 &   & 7 \\
\hline
U & 0.1 & Global & 1 & 5 &   & 15 \\
U & 0.1 & Tropics & 14 & 13 &   &   \\
U & 0.1 & SH extratropics & 5 & 3 & 1 & 12 \\
U & 0.1 & NH extratropics &   & 6 & 5 & 22 \\
U & 0.1 & All extratropics &   & 3 & 1 & 20 \\
\hline
U & 3 & Global & 4 & 10 & 1 & 8 \\
U & 3 & Tropics & 14 & 20 & 3 &   \\
U & 3 & SH extratropics & 11 & 7 & 3 & 4 \\
U & 3 & NH extratropics & 6 & 7 & 4 & 16 \\
U & 3 & All extratropics & 2 & 7 &   & 12 \\
\hline
\multicolumn{7}{l}{As Table~\ref{tab:rmsd_num_months}, but using \change{12 55$\frac{2}{3}$-sol}{30 $\sim$55.7-sol} months instead of \change{16 41.75-sol}{40 $\sim$41.8-sol} months.}\\ %
\end{tabular}
\label{tab:rmsd_num_months_12} %
\end{table}
\begin{table}[b]
\vspace{-.3in}
\caption{Information about probabilities of obtaining the observed results under various null hypotheses}
\begin{tabular}{|lll|l|l|r|l|}
\hline
      &   &   &   & Control-- &   &   \\
Field & Domain & Meridional & Reanalyses & reanalysis & $S_{obs}$ & More \\
      & top (hPa) & domain & not & differences &   & extreme \\
      &   &   & converging & same &   & $S$ \\
\hline
T & 0.1 & Global & $\mathbf{9.31 \times 10^{-10}}$ & $\mathit{3.25 \times 10^{-4}}$ & 12 & $\mathit{4.44 \times 10^{-4}}$ \\
T & 0.1 & Tropics & $\mathbf{2.89 \times 10^{-8}}$ & $\mathbf{8.43 \times 10^{-6}}$ & 5 & $6.24 \times 10^{-2}$ \\
T & 0.1 & SH extratropics & $\mathbf{9.31 \times 10^{-10}}$ & $\mathit{5.22 \times 10^{-3}}$ & 7 & $3.89 \times 10^{-2}$ \\
T & 0.1 & NH extratropics & $\mathbf{9.31 \times 10^{-10}}$ & $2.00 \times 10^{-1}$ & 12 & $1.69 \times 10^{-2}$ \\
T & 0.1 & All extratropics & $\mathbf{9.31 \times 10^{-10}}$ & $1.61 \times 10^{-2}$ & 11 & $1.25 \times 10^{-2}$ \\
\hline
T & 3 & Global & $\mathbf{9.31 \times 10^{-10}}$ & $\mathit{5.22 \times 10^{-3}}$ & 6 & $3.14 \times 10^{-2}$ \\
T & 3 & Tropics & $\mathbf{4.34 \times 10^{-7}}$ & $\mathbf{5.77 \times 10^{-8}}$ & 2 & $5.00 \times 10^{-1}$ \\
T & 3 & SH extratropics & $\mathbf{2.97 \times 10^{-5}}$ & $4.28 \times 10^{-2}$ & 5 & $1.80 \times 10^{-1}$ \\
T & 3 & NH extratropics & $\mathbf{9.31 \times 10^{-10}}$ & $9.87 \times 10^{-2}$ & 3 & $6.30 \times 10^{-1}$ \\
T & 3 & All extratropics & $\mathbf{9.31 \times 10^{-10}}$ & $1.61 \times 10^{-2}$ & 7 & $1.56 \times 10^{-2}$ \\
\hline
U & 0.1 & Global & $\mathbf{2.89 \times 10^{-8}}$ & $\mathit{3.25 \times 10^{-4}}$ & 15 & $\mathbf{5.90 \times 10^{-5}}$ \\
U & 0.1 & Tropics & $4.28 \times 10^{-1}$ & $5.85 \times 10^{-1}$ & 0 & $1$ \\
U & 0.1 & SH extratropics & $\mathit{1.62 \times 10^{-4}}$ & $\mathbf{8.43 \times 10^{-6}}$ & 11 & $\mathit{3.45 \times 10^{-3}}$ \\
U & 0.1 & NH extratropics & $\mathbf{9.31 \times 10^{-10}}$ & $\mathit{1.43 \times 10^{-3}}$ & 17 & $\mathit{4.31 \times 10^{-4}}$ \\
U & 0.1 & All extratropics & $\mathbf{9.31 \times 10^{-10}}$ & $\mathbf{8.43 \times 10^{-6}}$ & 19 & $\mathbf{1.70 \times 10^{-5}}$ \\
\hline
U & 3 & Global & $\mathbf{2.97 \times 10^{-5}}$ & $9.87 \times 10^{-2}$ & 7 & $3.95 \times 10^{-2}$ \\
U & 3 & Tropics & $4.28 \times 10^{-1}$ & $9.87 \times 10^{-2}$ & -3 & $2.50 \times 10^{-1}$ \\
U & 3 & SH extratropics & $1.00 \times 10^{-1}$ & $\mathit{5.22 \times 10^{-3}}$ & 1 & $1$ \\
U & 3 & NH extratropics & $\mathit{7.15 \times 10^{-4}}$ & $\mathit{5.22 \times 10^{-3}}$ & 12 & $1.17 \times 10^{-2}$ \\
U & 3 & All extratropics & $\mathbf{4.34 \times 10^{-7}}$ & $\mathit{5.22 \times 10^{-3}}$ & 12 & $\mathit{5.07 \times 10^{-4}}$ \\
\hline
%
\multicolumn{7}{l}{As Table~\ref{tab:pvals_16}, but using 30 $\sim$55.7-sol months.  This table should be used to help}\\
\multicolumn{7}{l}{interpret the results given in Table~\ref{tab:rmsd_num_months_12}.}\\
\end{tabular}
\label{tab:pvals_12} %
\end{table}

\section{Statistical analyses of RMS difference results}
\label{app:stats}

The arguments about reanalysis convergence and the relative sizes of the MCTRL and ECTRL biases made in section~\ref{sec:meanflow_convergence} are based on qualitative interpretation of Table\add{s}~\ref{tab:rmsd_num_months} \add{and~}\ref{tab:rmsd_num_months_12} and physical reasoning.  It is therefore worth investigating quantitatively how likely we are to have obtained these results under some relevant null hypotheses---could the apparent signals really just be internal variability noise? %

Let us first consider the apparent convergence of the UK-LMD and GFDL Mars general circulation model (MGCM) mean states when temperature data are assimilated (Table\add{s}~\ref{tab:rmsd_num_months}\add{ and~}\ref{tab:rmsd_num_months_12}, ``$rmsd(M_C,E_C)<rmsd(M_R,E_R)$'' column\add{s}).  We will assume (implausibly) that assimilating temperature data has no effect whatsoever on the monthly mean states of the MGCMs.  If this is so, then the MACDA--EMARS RMS differences should be drawn from the same probability density functions as the MCTRL--ECTRL RMS differences and for any given month both data set pairs should have an equal probability of having the smaller RMS difference. %

We will further postulate that the values of $rmsd(M_C,E_C)$ and $rmsd(M_R,E_R)$ for individual months are independent.  This assumption seems reasonable, as Martian atmospheric variability that has timescales longer than our \change{41.75}{$\sim$41.8}-sol months and that is not strongly radiatively forced by the annual cycle or via coupling to the dust field is apparently rare [e.g., {\it Banfield et al.}, 2004].  (The last qualifier is important because the dust fields in all four data sets are being constrained by observations and therefore we are interested only in forms of variability compatible with the prescribed dust fields.)  Given this postulate, it is easy to see that (under our null hypothesis of no data assimilation effect) the number of months for which $rmsd(M_C,E_C)<rmsd(M_R,E_R)$ is satisfied is drawn from a binomial distribution with a success probability of 0.5 \cite{wilks_ch4}. %

The probability of $rmsd(M_C,E_C)<rmsd(M_R,E_R)$ being satisfied for a number of months {\it less than or equal to} that actually observed is often quite small under the null hypothesis (Table\add{s}~\ref{tab:pvals_16}\add{ and~}\ref{tab:pvals_12}, ``reanalyses not converging'' column\add{s}).  In conjunction with the physical knowledge that data assimilation does in fact affect the MACDA and EMARS states, we conclude that assimilation of temperature retrievals into the MGCMs is bringing their monthly mean states closer together.  It seems unlikely that this result is solely due to data assimilation synchronizing the instantaneous weather states of models with the same underlying climate---this is because the (time-varying) weather should have been largely removed by taking the monthly means prior to computing the RMS differences.  We thus conclude that data assimilation is converging distinct MGCM climates. %

\begin{table}[b]
\vspace{-.3in}
\caption{Information about probabilities of obtaining the observed results under various null hypotheses}
\begin{tabular}{|lll|l|l|r|l|}
\hline
      &   &   &   & Control-- &   &   \\
Field & Domain & Meridional & Reanalyses & reanalysis & $S_{obs}$ & More \\
      & top (hPa) & domain & not & differences &   & extreme \\
      &   &   & converging & same &   & $S$ \\
\hline
T & 0.1 & Global & $\mathbf{9.09 \times 10^{-13}}$ & $\mathbf{4.23 \times 10^{-5}}$ & 14 & $\mathit{1.14 \times 10^{-4}}$ \\
T & 0.1 & Tropics & $\mathbf{3.73 \times 10^{-11}}$ & $\mathbf{1.38 \times 10^{-6}}$ & 6 & $3.13 \times 10^{-2}$ \\
T & 0.1 & SH extratropics & $\mathbf{9.09 \times 10^{-13}}$ & $1.66 \times 10^{-2}$ & 9 & $2.26 \times 10^{-2}$ \\
T & 0.1 & NH extratropics & $\mathbf{9.09 \times 10^{-13}}$ & $1.66 \times 10^{-2}$ & 17 & $\mathit{3.32 \times 10^{-3}}$ \\
T & 0.1 & All extratropics & $\mathbf{9.09 \times 10^{-13}}$ & $\mathit{1.82 \times 10^{-4}}$ & 15 & $\mathit{1.46 \times 10^{-3}}$ \\
\hline
T & 3 & Global & $\mathbf{9.09 \times 10^{-13}}$ & $\mathit{6.43 \times 10^{-3}}$ & 5 & $1.80 \times 10^{-1}$ \\
T & 3 & Tropics & $\mathbf{9.73 \times 10^{-9}}$ & $\mathbf{1.86 \times 10^{-7}}$ & 3 & $2.50 \times 10^{-1}$ \\
T & 3 & SH extratropics & $\mathbf{6.91 \times 10^{-7}}$ & $1.66 \times 10^{-2}$ & 5 & $2.27 \times 10^{-1}$ \\
T & 3 & NH extratropics & $\mathbf{9.09 \times 10^{-13}}$ & $1.54 \times 10^{-1}$ & 9 & $7.87 \times 10^{-2}$ \\
T & 3 & All extratropics & $\mathbf{9.09 \times 10^{-13}}$ & $1.66 \times 10^{-2}$ & 4 & $3.44 \times 10^{-1}$ \\
\hline
U & 0.1 & Global & $\mathbf{3.73 \times 10^{-11}}$ & $\mathit{2.22 \times 10^{-3}}$ & 19 & $\mathbf{3.00 \times 10^{-6}}$ \\
U & 0.1 & Tropics & $2.15 \times 10^{-1}$ & $6.36 \times 10^{-1}$ & 0 & $1$ \\
U & 0.1 & SH extratropics & $\mathbf{9.11 \times 10^{-5}}$ & $\mathbf{8.36 \times 10^{-6}}$ & 13 & $\mathit{1.02 \times 10^{-3}}$ \\
U & 0.1 & NH extratropics & $\mathbf{3.73 \times 10^{-11}}$ & $\mathit{6.80 \times 10^{-4}}$ & 22 & $\mathit{1.82 \times 10^{-4}}$ \\
U & 0.1 & All extratropics & $\mathbf{9.09 \times 10^{-13}}$ & $\mathbf{1.86 \times 10^{-7}}$ & 28 & $\mathbf{0}$ \\
\hline
U & 3 & Global & $\mathbf{9.29 \times 10^{-8}}$ & $2.68 \times 10^{-1}$ & 11 & $\mathit{7.55 \times 10^{-3}}$ \\
U & 3 & Tropics & $3.18 \times 10^{-1}$ & $8.07 \times 10^{-2}$ & -5 & $6.27 \times 10^{-2}$ \\
U & 3 & SH extratropics & $1.92 \times 10^{-2}$ & $\mathit{6.43 \times 10^{-3}}$ & 1 & $1$ \\
U & 3 & NH extratropics & $\mathbf{9.11 \times 10^{-5}}$ & $\mathit{6.43 \times 10^{-3}}$ & 14 & $\mathit{9.19 \times 10^{-3}}$ \\
U & 3 & All extratropics & $\mathbf{7.47 \times 10^{-10}}$ & $\mathit{6.43 \times 10^{-3}}$ & 14 & $\mathit{5.36 \times 10^{-4}}$ \\
\hline
\multicolumn{7}{l}{See text of \ref{app:stats} for further details, including descriptions of the columns.}\\
\multicolumn{7}{l}{Calculations were done using \change{16 41.75-sol}{40 $\sim$41.8-sol} months, and thus this table should be used to help}\\
\multicolumn{7}{l}{interpret the results given in Table~\ref{tab:rmsd_num_months}.  Probabilities $<10^{-4}$ are written}\\
\multicolumn{7}{l}{in {\bf bold}, while probabilities $<10^{-2}$ are {\it italicized}.}\\
\end{tabular}
\label{tab:pvals_16}
\end{table}

The first step in our argument that the EMARS control simulation is likely less biased than its MACDA counterpart is that the inequality $rmsd(M_R,M_C)<rmsd(E_R,E_C)$ is satisfied in only a minority of months for nearly all field--region combinations of interest.  %
Next we will compute whether these results could have been obtained if $rmsd(M_R,M_C)$ and $rmsd(E_R,E_C)$ are in fact drawn from the same probability density functions---one reasonable way to operationalize the null hypothesis that MCTRL and ECTRL agree equally well with their associated reanalyses.  %

Our analysis of this case parallels that used to investigate whether data assimilation brings the MGCMs' mean states together.  %
We see that under our null hypothesis that $rmsd(M_R,M_C)$ and $rmsd(E_R,E_C)$ are drawn from the same probability density functions, the number of months for which $rmsd(M_R,M_C)<rmsd(E_R,E_C)$ is satisfied is again drawn from a binomial distribution with a success probability of 0.5.  %
Under this null hypothesis, the probability of $rmsd(M_R,M_C)<rmsd(E_R,E_C)$ being satisfied for a number of months {\it as or more extreme} than actually observed is often fairly low (Table\add{s}~\ref{tab:pvals_16}\add{ and~}\ref{tab:pvals_12}, ``control--reanalysis differences same'' column\add{s}).  %
In other words, if \add{there are $N_{tot}$ months total and }$rmsd(M_R,M_C)<rmsd(E_R,E_C)$ \change{was}{is} actually satisfied in $N_{obs}$ \change{months}{of them} the listed value is the probability (under the null hypothesis) of it being satisfied in $N$ months, where $0\le N\le N_{obs}$ {\it or} \change{$(16-N_{obs})\le N\le 16$}{$(N_{tot}-N_{obs})\le N\le N_{tot}$}.  %
\add{($N_{tot}$ is of course 40 (30) for the $\sim$41.8-sol ($\sim$55.7-sol) month case.)}  %

We use this two-tailed statistical test because both very small {\it and} very large values of $N_{obs}$ are unlikely to be observed under the stated null hypothesis.  %
This contrasts with our use of an implicitly one-tailed test when examining whether data assimilation converges the MGCM states---a one-tailed test was appropriate in that case because satisfaction of $rmsd(M_C,E_C)<rmsd(M_R,E_R)$ in a large number of months would be inconsistent with the alternative hypothesis that data assimilation brings the MGCMs' mean states together.  %

The second step of our argument for smaller ECTRL biases involved comparing the rightmost two columns of Table\add{s}~\ref{tab:rmsd_num_months}\add{ and~}\ref{tab:rmsd_num_months_12}.  %
We noted that $rmsd(M_R,E_C)<rmsd(M_R,M_C)$ was \change{always}{generally} satisfied in at least as many months as $rmsd(E_R,M_C)<rmsd(E_R,E_C)$.  %
Let us define a test statistic $S$, where $S$ is the number of months in which $rmsd(M_R,E_C)<rmsd(M_R,M_C)$ was satisfied minus the number of months in which $rmsd(E_R,M_C)<rmsd(E_R,E_C)$.  %
Further denoting an observed value of $S$ as $S_{obs}$, we essentially argued that ECTRL was less biased because we \change{always}{usually} found $S_{obs}\ge 0$.  %

The null hypothesis we will evaluate in this case is that the ECTRL and MCTRL simulations are simply different realizations of internal variability and that these versions of the free-running GFDL and UK-LMD MGCMs actually have the same underlying climate (given the imposed dust fields).  %
We thus postulate that the ECTRL and MCTRL monthly mean states are drawn from same (month-dependent) probability density functions, and also continue to assume that the monthly mean states for a given month are drawn independently of those for all other months.  %

If this null hypothesis is true, for each of \change{16}{the $N_{tot}$ total} months we are essentially drawing two monthly mean states from a (month-dependent) probability density function and randomly assigning the label ``ECTRL'' to one mean state and ``MCTRL'' to the other.  %
We can thus evaluate the null hypothesis using a permutation test \cite{wilks_ch5}: for each of the \change{16}{$N_{tot}$} months, we can independently choose to exchange (or not exchange) the ``ECTRL'' and ``MCTRL'' labels attached to the monthly mean states.  %
\change{This yields $2^{16}=65536$}{There are thus $2^{N_{tot}}$ possible} distinct synthetic \change{sets of}{labelings of the} ECTRL and MCTRL monthly mean states.  %
Exactly one of these \change{sets}{labelings} (the one without any \remove{label }exchanges) matches the actual ECTRL and MCTRL states, but if the null hypothesis is true {\it we are equally likely to have observed any of these \change{sets}{labelings}}.  %

For each field--region combination of interest, we can thus use \change{the $2^{16}$}{these} synthetic \change{sets of}{labelings of the} monthly mean states to compute the appropriate null distribution for $S$.  %
\add{In practice, generating all $2^{N_{tot}}$ ($>$$10^9$ even for $N_{tot}=30$) synthetic sets is computationally intractable---we therefore approximate the $S$ null distribution by drawing $10^6$ of the sets at random.}
We then calculate $S_{obs}$ values (Table\add{s}~\ref{tab:pvals_16}\add{ and~}\ref{tab:pvals_12}, ``$S_{obs}$'' column\add{s}) and use the \add{approximate }null distributions to determine the probability of obtaining an $S$ value {\it as or more extreme} than actually observed.  %
By ``as or more extreme'' we mean %
$|S|\ge |S_{obs}|$---we are thus conducting a two-tailed test, as both large and small values of $S_{obs}$ would argue against our chosen null hypothesis.  %
Our results are shown in the rightmost column\add{s} of Table\add{s}~\ref{tab:pvals_16}\add{ and~}\ref{tab:pvals_12} (``more extreme $S$'').  %

Although for some field--region combinations the $S_{obs}$ value is found to be fully consistent with the null hypothesis, in \add{most }cases with $p_t~=~0.1$ hPa the probability of getting an $S$ value at least as extreme as observed is substantially less than 1.  %
In conjunction with the known structural differences between the two MGCMs, this finding further supports the idea that the UK-LMD and GFDL MGCMs do in fact have different climates and that the apparent superiority of ECTRL over MCTRL is not simply a random manifestation of internal variability.  %

\acknowledgments
We particularly thank Tiffany A. Shaw for her support of this work---she provided much helpful input on early versions of this paper and financially supported TAM and GED.  TAM was funded through a fellowship from the David and Lucile Packard Foundation, while GED was funded through National Science Foundation grant AGS-1742944 via Leadership Alliance.  GED also acknowledges participation in the Leadership Alliance Summer Research Early Identification Program at the University of Chicago.  SJG is supported by NASA Mars Data Analysis Program grant 80NSSC17K0690.  Reduced data necessary to reproduce all figures and tables of this paper are available from Knowledge@UChicago [get link after depositing final data].  The full MACDA reanalysis data set may be downloaded from the British Atmospheric Data Centre\add{ at }\url{https://doi.org/10.5285/78114093-E2BD-4601-8AE5-3551E62AEF2B}, while the full MACDA control simulation is available upon request from Luca Montabone \cite{montabone14}.  The full EMARS reanalysis and \remove{part of the EMARS }control simulation \remove{used in this paper }may be downloaded from the Penn State Data Commons \add{at }\url{https://doi.org/10.18113/D3W375}.  Finally, we thank Pragallva Barpanda and R. John Wilson for comments on a draft of this paper.

\bibliography{mars_momentum_2019}

\begin{thebibliography}{}

\bibitem [\protect \citeauthoryear {%
Banfield%
, Conrath%
, Gierasch%
, Wilson%
\BCBL {}\ \BBA {} Smith%
}{%
Banfield%
\ \protect \BOthers {.}}{%
{\protect \APACyear {2004}}%
}]{%
banfield04}
\APACinsertmetastar {%
banfield04}%
\begin{APACrefauthors}%
Banfield, D.%
, Conrath, B\BPBI J.%
, Gierasch, P\BPBI J.%
, Wilson, R\BPBI J.%
\BCBL {}\ \BBA {} Smith, M\BPBI D.%
\end{APACrefauthors}%
\unskip\
\newblock
\APACrefYearMonthDay{2004}{}{}.
\newblock
{\BBOQ}\APACrefatitle {Traveling waves in the {M}artian atmosphere from {MGS
  TES} Nadir data} {Traveling waves in the {M}artian atmosphere from {MGS TES}
  nadir data}.{\BBCQ}
\newblock
\APACjournalVolNumPages{Icarus}{170}{2}{365-403}.
\newblock
\begin{APACrefDOI} \doi{10.1016/j.icarus.2004.03.015} \end{APACrefDOI}
\PrintBackRefs{\CurrentBib}

\bibitem [\protect \citeauthoryear {%
Clancy%
\ \protect \BOthers {.}}{%
Clancy%
\ \protect \BOthers {.}}{%
{\protect \APACyear {2000}}%
}]{%
clancy00}
\APACinsertmetastar {%
clancy00}%
\begin{APACrefauthors}%
Clancy, R\BPBI T.%
, Sandor, B\BPBI J.%
, Wolff, M\BPBI J.%
, Christensen, P\BPBI R.%
, Smith, M\BPBI D.%
, Pearl, J\BPBI C.%
\BDBL {}Wilson, R\BPBI J.%
\end{APACrefauthors}%
\unskip\
\newblock
\APACrefYearMonthDay{2000}{}{}.
\newblock
{\BBOQ}\APACrefatitle {An intercomparison of ground-based millimeter, {MGS
  TES}, and {V}iking atmospheric temperature measurements: Seasonal and
  interannual variability of temperatures and dust loading in the global {M}ars
  atmosphere} {An intercomparison of ground-based millimeter, {MGS TES}, and
  {V}iking atmospheric temperature measurements: Seasonal and interannual
  variability of temperatures and dust loading in the global {M}ars
  atmosphere}.{\BBCQ}
\newblock
\APACjournalVolNumPages{J. Geophys. Res.}{105}{E4}{9553-9571}.
\newblock
\begin{APACrefDOI} \doi{10.1029/1999JE001089} \end{APACrefDOI}
\PrintBackRefs{\CurrentBib}

\bibitem [\protect \citeauthoryear {%
Conrath%
}{%
Conrath%
}{%
{\protect \APACyear {1975}}%
}]{%
conrath75}
\APACinsertmetastar {%
conrath75}%
\begin{APACrefauthors}%
Conrath, B\BPBI J.%
\end{APACrefauthors}%
\unskip\
\newblock
\APACrefYearMonthDay{1975}{}{}.
\newblock
{\BBOQ}\APACrefatitle {Thermal structure of the {M}artian atmosphere during the
  dissipation of the dust storm of 1971} {Thermal structure of the {M}artian
  atmosphere during the dissipation of the dust storm of 1971}.{\BBCQ}
\newblock
\APACjournalVolNumPages{Icarus}{24}{1}{36-46}.
\newblock
\begin{APACrefDOI} \doi{10.1016/0019-1035(75)90156-6} \end{APACrefDOI}
\PrintBackRefs{\CurrentBib}

\bibitem [\protect \citeauthoryear {%
Conrath%
\ \protect \BOthers {.}}{%
Conrath%
\ \protect \BOthers {.}}{%
{\protect \APACyear {2000}}%
}]{%
conrath00}
\APACinsertmetastar {%
conrath00}%
\begin{APACrefauthors}%
Conrath, B\BPBI J.%
, Pearl, J\BPBI C.%
, Smith, M\BPBI D.%
, Maguire, W\BPBI C.%
, Christensen, P\BPBI R.%
, Dason, S.%
\BCBL {}\ \BBA {} Kaelberer, M\BPBI S.%
\end{APACrefauthors}%
\unskip\
\newblock
\APACrefYearMonthDay{2000}{}{}.
\newblock
{\BBOQ}\APACrefatitle {{Mars Global Surveyor Thermal Emission Spectrometer
  (TES)} observations: Atmospheric temperatures during aerobraking and science
  phasing} {{Mars Global Surveyor Thermal Emission Spectrometer (TES)}
  observations: Atmospheric temperatures during aerobraking and science
  phasing}.{\BBCQ}
\newblock
\APACjournalVolNumPages{J. Geophys. Res.}{105}{E4}{9509-9519}.
\newblock
\begin{APACrefDOI} \doi{10.1029/1999JE001095} \end{APACrefDOI}
\PrintBackRefs{\CurrentBib}

\bibitem [\protect \citeauthoryear {%
Cremons%
\ \protect \BOthers {.}}{%
Cremons%
\ \protect \BOthers {.}}{%
{\protect \APACyear {2020}}%
}]{%
cremons20}
\APACinsertmetastar {%
cremons20}%
\begin{APACrefauthors}%
Cremons, D\BPBI R.%
, Abshire, J\BPBI B.%
, Sun, X.%
, Allan, G.%
, Riris, H.%
, Smith, M\BPBI D.%
\BDBL {}Hovis, F.%
\end{APACrefauthors}%
\unskip\
\newblock
\APACrefYearMonthDay{2020}{}{}.
\newblock
{\BBOQ}\APACrefatitle {Design of a direct-detection wind and aerosol lidar for
  mars orbit} {Design of a direct-detection wind and aerosol lidar for mars
  orbit}.{\BBCQ}
\newblock
\APACjournalVolNumPages{{CEAS} Space J.}{12}{2}{149-162}.
\newblock
\begin{APACrefDOI} \doi{10.1007/s12567-020-00301-z} \end{APACrefDOI}
\PrintBackRefs{\CurrentBib}

\bibitem [\protect \citeauthoryear {%
Forget%
\ \protect \BOthers {.}}{%
Forget%
\ \protect \BOthers {.}}{%
{\protect \APACyear {1999}}%
}]{%
forget99}
\APACinsertmetastar {%
forget99}%
\begin{APACrefauthors}%
Forget, F.%
, Hourdin, F.%
, Fournier, R.%
, Hourdin, C.%
, Talagrand, O.%
, Collins, M.%
\BDBL {}Huot, J\BHBI P.%
\end{APACrefauthors}%
\unskip\
\newblock
\APACrefYearMonthDay{1999}{}{}.
\newblock
{\BBOQ}\APACrefatitle {Improved general circulation models of the {M}artian
  atmosphere from the surface to above 80 km} {Improved general circulation
  models of the {M}artian atmosphere from the surface to above 80 km}.{\BBCQ}
\newblock
\APACjournalVolNumPages{J. Geophys. Res.}{104}{E10}{24155-24175}.
\newblock
\begin{APACrefDOI} \doi{10.1029/1999JE001025} \end{APACrefDOI}
\PrintBackRefs{\CurrentBib}

\bibitem [\protect \citeauthoryear {%
Gelaro%
\ \protect \BOthers {.}}{%
Gelaro%
\ \protect \BOthers {.}}{%
{\protect \APACyear {2017}}%
}]{%
gelaro17}
\APACinsertmetastar {%
gelaro17}%
\begin{APACrefauthors}%
Gelaro, R.%
, McCarty, W.%
, Su\'arez, M\BPBI J.%
, Todling, R.%
, Molod, A.%
, Takacs, L.%
\BDBL {}Zhao, B.%
\end{APACrefauthors}%
\unskip\
\newblock
\APACrefYearMonthDay{2017}{}{}.
\newblock
{\BBOQ}\APACrefatitle {The {Modern-Era Retrospective Analysis for Research and
  Applications}, Version 2 {(MERRA-2)}} {The {Modern-Era Retrospective Analysis
  for Research and Applications}, version 2 {(MERRA-2)}}.{\BBCQ}
\newblock
\APACjournalVolNumPages{J. Climate}{30}{14}{5419-5454}.
\newblock
\begin{APACrefDOI} \doi{10.1175/JCLI-D-16-0758.1} \end{APACrefDOI}
\PrintBackRefs{\CurrentBib}

\bibitem [\protect \citeauthoryear {%
Greybush%
, Gillespie%
\BCBL {}\ \BBA {} Wilson%
}{%
Greybush%
, Gillespie%
\BCBL {}\ \BBA {} Wilson%
}{%
{\protect \APACyear {2019}}%
}]{%
greybush19te}
\APACinsertmetastar {%
greybush19te}%
\begin{APACrefauthors}%
Greybush, S\BPBI J.%
, Gillespie, H\BPBI E.%
\BCBL {}\ \BBA {} Wilson, R\BPBI J.%
\end{APACrefauthors}%
\unskip\
\newblock
\APACrefYearMonthDay{2019}{}{}.
\newblock
{\BBOQ}\APACrefatitle {{Transient eddies in the TES/MCS Ensemble Mars
  Atmosphere Reanalysis System (EMARS)}} {{Transient eddies in the TES/MCS
  Ensemble Mars Atmosphere Reanalysis System (EMARS)}}.{\BBCQ}
\newblock
\APACjournalVolNumPages{Icarus}{317}{}{158-181}.
\newblock
\begin{APACrefDOI} \doi{10.1016/j.icarus.2018.07.001} \end{APACrefDOI}
\PrintBackRefs{\CurrentBib}

\bibitem [\protect \citeauthoryear {%
Greybush%
, Kalnay%
\BCBL {}\ \protect \BOthers {.}}{%
Greybush%
, Kalnay%
\BCBL {}\ \protect \BOthers {.}}{%
{\protect \APACyear {2019}}%
}]{%
greybush19gdj}
\APACinsertmetastar {%
greybush19gdj}%
\begin{APACrefauthors}%
Greybush, S\BPBI J.%
, Kalnay, E.%
, Wilson, R\BPBI J.%
, Hoffman, R\BPBI N.%
, Nehrkorn, T.%
, Leidner, M.%
\BDBL {}Miyoshi, T.%
\end{APACrefauthors}%
\unskip\
\newblock
\APACrefYearMonthDay{2019}{}{}.
\newblock
{\BBOQ}\APACrefatitle {The {Ensemble Mars Atmosphere Reanalysis System (EMARS)}
  Version 1.0} {The {Ensemble Mars Atmosphere Reanalysis System (EMARS)}
  version 1.0}.{\BBCQ}
\newblock
\APACjournalVolNumPages{Geosci. Data J.}{6}{2}{137-150}.
\newblock
\begin{APACrefDOI} \doi{10.1002/gdj3.77} \end{APACrefDOI}
\PrintBackRefs{\CurrentBib}

\bibitem [\protect \citeauthoryear {%
Harri%
\ \protect \BOthers {.}}{%
Harri%
\ \protect \BOthers {.}}{%
{\protect \APACyear {2017}}%
}]{%
harri17}
\APACinsertmetastar {%
harri17}%
\begin{APACrefauthors}%
Harri, A\BHBI M.%
, Pichkadze, K.%
, Zeleny, L.%
, Vazquez, L.%
, Schmidt, W.%
, Alexashkin, S.%
\BDBL {}Romero, P.%
\end{APACrefauthors}%
\unskip\
\newblock
\APACrefYearMonthDay{2017}{}{}.
\newblock
{\BBOQ}\APACrefatitle {The {MetNet} vehicle: a lander to deploy environmental
  stations for local and global investigations of {Mars}} {The {MetNet}
  vehicle: a lander to deploy environmental stations for local and global
  investigations of {Mars}}.{\BBCQ}
\newblock
\APACjournalVolNumPages{Geoscientific Instrumentation, Methods and Data
  Systems}{6}{1}{103--124}.
\newblock
\begin{APACrefDOI} \doi{10.5194/gi-6-103-2017} \end{APACrefDOI}
\PrintBackRefs{\CurrentBib}

\bibitem [\protect \citeauthoryear {%
Hersbach%
\ \protect \BOthers {.}}{%
Hersbach%
\ \protect \BOthers {.}}{%
{\protect \APACyear {2020}}%
}]{%
hersbach20}
\APACinsertmetastar {%
hersbach20}%
\begin{APACrefauthors}%
Hersbach, H.%
, Bell, B.%
, Berrisford, P.%
, Hirahara, S.%
, Hor\'anyi, A.%
, Mu\~noz\ Sabater, J.%
\BDBL {}Th\'epaut, J\BHBI N.%
\end{APACrefauthors}%
\unskip\
\newblock
\APACrefYearMonthDay{2020}{}{}.
\newblock
{\BBOQ}\APACrefatitle {The {ERA5} global reanalysis} {The {ERA5} global
  reanalysis}.{\BBCQ}
\newblock
\APACjournalVolNumPages{Q. J. Roy. Meteor. Soc.}{146}{730}{1999-2049}.
\newblock
\begin{APACrefDOI} \doi{10.1002/qj.3803} \end{APACrefDOI}
\PrintBackRefs{\CurrentBib}

\bibitem [\protect \citeauthoryear {%
Hinson%
}{%
Hinson%
}{%
{\protect \APACyear {2008}}%
}]{%
hinson08}
\APACinsertmetastar {%
hinson08}%
\begin{APACrefauthors}%
Hinson, D\BPBI P.%
\end{APACrefauthors}%
\unskip\
\newblock
\APACrefYearMonthDay{2008}{}{}.
\newblock
\APACrefbtitle {{Mars Global Surveyor} Radio Occultation Profiles of the
  Neutral Atmosphere - Reorganized} {{Mars Global Surveyor} radio occultation
  profiles of the neutral atmosphere - reorganized}\ (\BVOL\
  USA\_NASA\_JPL\_MORS\_1101).
\newblock
\APACrefnote{{NASA Planetary Data System}, MGS-M-RSS-5-TPS-V1.0, see figure
  available at
  \texttt{https://atmos.nmsu.edu/data\_and\_services/atmospheres\_data/MARS/tp.html},
  accessed June 18, 2019}
\PrintBackRefs{\CurrentBib}

\bibitem [\protect \citeauthoryear {%
Hinson%
, Simpson%
, Twicken%
, Tyler%
\BCBL {}\ \BBA {} Flasar%
}{%
Hinson%
\ \protect \BOthers {.}}{%
{\protect \APACyear {1999}}%
}]{%
hinson99}
\APACinsertmetastar {%
hinson99}%
\begin{APACrefauthors}%
Hinson, D\BPBI P.%
, Simpson, R\BPBI A.%
, Twicken, J\BPBI D.%
, Tyler, G\BPBI L.%
\BCBL {}\ \BBA {} Flasar, F\BPBI M.%
\end{APACrefauthors}%
\unskip\
\newblock
\APACrefYearMonthDay{1999}{}{}.
\newblock
{\BBOQ}\APACrefatitle {Initial results from radio occultation measurements with
  {M}ars {G}lobal {S}urveyor} {Initial results from radio occultation
  measurements with {M}ars {G}lobal {S}urveyor}.{\BBCQ}
\newblock
\APACjournalVolNumPages{J. Geophys. Res.}{104}{E11}{26997-27012}.
\newblock
\begin{APACrefDOI} \doi{10.1029/1999JE001069} \end{APACrefDOI}
\PrintBackRefs{\CurrentBib}

\bibitem [\protect \citeauthoryear {%
Hinson%
\ \BBA {} Wilson%
}{%
Hinson%
\ \BBA {} Wilson%
}{%
{\protect \APACyear {2004}}%
}]{%
hinson04}
\APACinsertmetastar {%
hinson04}%
\begin{APACrefauthors}%
Hinson, D\BPBI P.%
\BCBT {}\ \BBA {} Wilson, R\BPBI J.%
\end{APACrefauthors}%
\unskip\
\newblock
\APACrefYearMonthDay{2004}{}{}.
\newblock
{\BBOQ}\APACrefatitle {Temperature inversions, thermal tides, and water ice
  clouds in the Martian tropics} {Temperature inversions, thermal tides, and
  water ice clouds in the martian tropics}.{\BBCQ}
\newblock
\APACjournalVolNumPages{Journal of Geophysical Research:
  Planets}{109}{E1}{E01002}.
\newblock
\begin{APACrefDOI} \doi{10.1029/2003JE002129} \end{APACrefDOI}
\PrintBackRefs{\CurrentBib}

\bibitem [\protect \citeauthoryear {%
Hoffman%
\ \protect \BOthers {.}}{%
Hoffman%
\ \protect \BOthers {.}}{%
{\protect \APACyear {2010}}%
}]{%
hoffman10}
\APACinsertmetastar {%
hoffman10}%
\begin{APACrefauthors}%
Hoffman, M\BPBI J.%
, Greybush, S\BPBI J.%
, Wilson, R\BPBI J.%
, Gyarmati, G.%
, Hoffman, R\BPBI N.%
, Kalnay, E.%
\BDBL {}Szunyogh, I.%
\end{APACrefauthors}%
\unskip\
\newblock
\APACrefYearMonthDay{2010}{}{}.
\newblock
{\BBOQ}\APACrefatitle {An ensemble {K}alman filter data assimilation system for
  the {M}artian atmosphere: Implementation and simulation experiments} {An
  ensemble {K}alman filter data assimilation system for the {M}artian
  atmosphere: Implementation and simulation experiments}.{\BBCQ}
\newblock
\APACjournalVolNumPages{Icarus}{209}{2}{470-481}.
\newblock
\begin{APACrefDOI} \doi{10.1016/j.icarus.2010.03.034} \end{APACrefDOI}
\PrintBackRefs{\CurrentBib}

\bibitem [\protect \citeauthoryear {%
Holmes%
, Lewis%
\BCBL {}\ \BBA {} Patel%
}{%
Holmes%
\ \protect \BOthers {.}}{%
{\protect \APACyear {2020}}%
}]{%
holmes20}
\APACinsertmetastar {%
holmes20}%
\begin{APACrefauthors}%
Holmes, J\BPBI A.%
, Lewis, S\BPBI R.%
\BCBL {}\ \BBA {} Patel, M\BPBI R.%
\end{APACrefauthors}%
\unskip\
\newblock
\APACrefYearMonthDay{2020}{}{}.
\newblock
{\BBOQ}\APACrefatitle {{OpenMARS}: A global record of martian weather from 1999
  to 2015} {{OpenMARS}: A global record of martian weather from 1999 to
  2015}.{\BBCQ}
\newblock
\APACjournalVolNumPages{Planetary and Space Science}{188}{}{104962}.
\newblock
\begin{APACrefDOI} \doi{10.1016/j.pss.2020.104962} \end{APACrefDOI}
\PrintBackRefs{\CurrentBib}

\bibitem [\protect \citeauthoryear {%
Holmes%
, Lewis%
, Patel%
\BCBL {}\ \BBA {} Lef\`{e}vre%
}{%
Holmes%
\ \protect \BOthers {.}}{%
{\protect \APACyear {2018}}%
}]{%
holmes18}
\APACinsertmetastar {%
holmes18}%
\begin{APACrefauthors}%
Holmes, J\BPBI A.%
, Lewis, S\BPBI R.%
, Patel, M\BPBI R.%
\BCBL {}\ \BBA {} Lef\`{e}vre, F.%
\end{APACrefauthors}%
\unskip\
\newblock
\APACrefYearMonthDay{2018}{}{}.
\newblock
{\BBOQ}\APACrefatitle {A reanalysis of ozone on {M}ars from assimilation of
  {SPICAM} observations} {A reanalysis of ozone on {M}ars from assimilation of
  {SPICAM} observations}.{\BBCQ}
\newblock
\APACjournalVolNumPages{Icarus}{302}{}{308 - 318}.
\newblock
\begin{APACrefDOI} \doi{10.1016/j.icarus.2017.11.026} \end{APACrefDOI}
\PrintBackRefs{\CurrentBib}

\bibitem [\protect \citeauthoryear {%
Holmes%
, Lewis%
, Patel%
\BCBL {}\ \BBA {} Smith%
}{%
Holmes%
\ \protect \BOthers {.}}{%
{\protect \APACyear {2019}}%
}]{%
holmes19}
\APACinsertmetastar {%
holmes19}%
\begin{APACrefauthors}%
Holmes, J\BPBI A.%
, Lewis, S\BPBI R.%
, Patel, M\BPBI R.%
\BCBL {}\ \BBA {} Smith, M\BPBI D.%
\end{APACrefauthors}%
\unskip\
\newblock
\APACrefYearMonthDay{2019}{}{}.
\newblock
{\BBOQ}\APACrefatitle {Global analysis and forecasts of carbon monoxide on
  {Mars}} {Global analysis and forecasts of carbon monoxide on {Mars}}.{\BBCQ}
\newblock
\APACjournalVolNumPages{Icarus}{328}{}{232 - 245}.
\newblock
\begin{APACrefDOI} \doi{10.1016/j.icarus.2019.03.016} \end{APACrefDOI}
\PrintBackRefs{\CurrentBib}

\bibitem [\protect \citeauthoryear {%
Houben%
}{%
Houben%
}{%
{\protect \APACyear {1999}}%
}]{%
houben99}
\APACinsertmetastar {%
houben99}%
\begin{APACrefauthors}%
Houben, H.%
\end{APACrefauthors}%
\unskip\
\newblock
\APACrefYearMonthDay{1999}{}{}.
\newblock
{\BBOQ}\APACrefatitle {Assimilation of {Mars Global Surveyor} meteorological
  data} {Assimilation of {Mars Global Surveyor} meteorological data}.{\BBCQ}
\newblock
\APACjournalVolNumPages{Advances in Space Research}{23}{11}{1899 - 1902}.
\newblock
\begin{APACrefDOI} \doi{10.1016/S0273-1177(99)00273-2} \end{APACrefDOI}
\PrintBackRefs{\CurrentBib}

\bibitem [\protect \citeauthoryear {%
Kursinski%
, McCormick%
\BCBL {}\ \BBA {} Folkner%
}{%
Kursinski%
\ \protect \BOthers {.}}{%
{\protect \APACyear {2012}}%
}]{%
kursinski12}
\APACinsertmetastar {%
kursinski12}%
\begin{APACrefauthors}%
Kursinski, E\BPBI R.%
, McCormick, C\BPBI C.%
\BCBL {}\ \BBA {} Folkner, W\BPBI M.%
\end{APACrefauthors}%
\unskip\
\newblock
\APACrefYearMonthDay{2012}{}{}.
\newblock
\APACrefbtitle {An orbiting {M}ars atmosphere, gravity, navigation and
  telecommunications system.} {An orbiting {M}ars atmosphere, gravity,
  navigation and telecommunications system.}
\newblock
\APACrefnote{Paper presented at Concepts and Approaches for Mars Exploration,
  USRA, Houston, Texas, 12--14 June, available electronically at
  \texttt{https://www.lpi.usra.edu/meetings/marsconcepts2012/pdf/4357.pdf},
  accessed July 24, 2019}
\PrintBackRefs{\CurrentBib}

\bibitem [\protect \citeauthoryear {%
Lee%
\ \protect \BOthers {.}}{%
Lee%
\ \protect \BOthers {.}}{%
{\protect \APACyear {2011}}%
}]{%
lee11}
\APACinsertmetastar {%
lee11}%
\begin{APACrefauthors}%
Lee, C.%
, Lawson, W\BPBI G.%
, Richardson, M\BPBI I.%
, Anderson, J\BPBI L.%
, Collins, N.%
, Hoar, T.%
\BCBL {}\ \BBA {} Mischna, M.%
\end{APACrefauthors}%
\unskip\
\newblock
\APACrefYearMonthDay{2011}{}{}.
\newblock
{\BBOQ}\APACrefatitle {Demonstration of ensemble data assimilation for {Mars}
  using {DART}, {MarsWRF}, and radiance observations from {MGS TES}}
  {Demonstration of ensemble data assimilation for {Mars} using {DART},
  {MarsWRF}, and radiance observations from {MGS TES}}.{\BBCQ}
\newblock
\APACjournalVolNumPages{Journal of Geophysical Research:
  Planets}{116}{E11}{E11011}.
\newblock
\begin{APACrefDOI} \doi{10.1029/2011JE003815} \end{APACrefDOI}
\PrintBackRefs{\CurrentBib}

\bibitem [\protect \citeauthoryear {%
Lewis%
, Collins%
\BCBL {}\ \BBA {} Read%
}{%
Lewis%
\ \protect \BOthers {.}}{%
{\protect \APACyear {1997}}%
}]{%
lewis97}
\APACinsertmetastar {%
lewis97}%
\begin{APACrefauthors}%
Lewis, S\BPBI R.%
, Collins, M.%
\BCBL {}\ \BBA {} Read, P\BPBI L.%
\end{APACrefauthors}%
\unskip\
\newblock
\APACrefYearMonthDay{1997}{}{}.
\newblock
{\BBOQ}\APACrefatitle {Data assimilation with a {M}artian atmospheric {GCM}: An
  example using thermal data} {Data assimilation with a {M}artian atmospheric
  {GCM}: An example using thermal data}.{\BBCQ}
\newblock
\APACjournalVolNumPages{Adv. Space Res.}{19}{8}{1267-1270}.
\newblock
\begin{APACrefDOI} \doi{10.1016/S0273-1177(97)00280-9} \end{APACrefDOI}
\PrintBackRefs{\CurrentBib}

\bibitem [\protect \citeauthoryear {%
Lewis%
\ \BBA {} Read%
}{%
Lewis%
\ \BBA {} Read%
}{%
{\protect \APACyear {1995}}%
}]{%
lewis95}
\APACinsertmetastar {%
lewis95}%
\begin{APACrefauthors}%
Lewis, S\BPBI R.%
\BCBT {}\ \BBA {} Read, P\BPBI L.%
\end{APACrefauthors}%
\unskip\
\newblock
\APACrefYearMonthDay{1995}{}{}.
\newblock
{\BBOQ}\APACrefatitle {An Operational Data Assimilation Scheme for the
  {M}artian Atmosphere} {An operational data assimilation scheme for the
  {M}artian atmosphere}.{\BBCQ}
\newblock
\APACjournalVolNumPages{Adv. Space Res.}{16}{6}{9-13}.
\newblock
\begin{APACrefDOI} \doi{10.1016/0273-1177(95)00244-9} \end{APACrefDOI}
\PrintBackRefs{\CurrentBib}

\bibitem [\protect \citeauthoryear {%
Lewis%
, Read%
\BCBL {}\ \BBA {} Collins%
}{%
Lewis%
\ \protect \BOthers {.}}{%
{\protect \APACyear {1996}}%
}]{%
lewis96}
\APACinsertmetastar {%
lewis96}%
\begin{APACrefauthors}%
Lewis, S\BPBI R.%
, Read, P\BPBI L.%
\BCBL {}\ \BBA {} Collins, M.%
\end{APACrefauthors}%
\unskip\
\newblock
\APACrefYearMonthDay{1996}{}{}.
\newblock
{\BBOQ}\APACrefatitle {Martian atmospheric data assimilation with a simplified
  general circulation model: Orbiter and lander networks} {Martian atmospheric
  data assimilation with a simplified general circulation model: Orbiter and
  lander networks}.{\BBCQ}
\newblock
\APACjournalVolNumPages{Planet. Space Sci.}{44}{11}{1395-1409}.
\newblock
\begin{APACrefDOI} \doi{10.1016/S0032-0633(96)00058-X} \end{APACrefDOI}
\PrintBackRefs{\CurrentBib}

\bibitem [\protect \citeauthoryear {%
Lewis%
, Read%
, Conrath%
, Pearl%
\BCBL {}\ \BBA {} Smith%
}{%
Lewis%
\ \protect \BOthers {.}}{%
{\protect \APACyear {2007}}%
}]{%
lewis07}
\APACinsertmetastar {%
lewis07}%
\begin{APACrefauthors}%
Lewis, S\BPBI R.%
, Read, P\BPBI L.%
, Conrath, B\BPBI J.%
, Pearl, J\BPBI C.%
\BCBL {}\ \BBA {} Smith, M\BPBI D.%
\end{APACrefauthors}%
\unskip\
\newblock
\APACrefYearMonthDay{2007}{}{}.
\newblock
{\BBOQ}\APACrefatitle {Assimilation of {T}hermal {E}mission {S}pectrometer
  atmospheric data during the {M}ars {G}lobal {S}urveyor aerobraking period}
  {Assimilation of {T}hermal {E}mission {S}pectrometer atmospheric data during
  the {M}ars {G}lobal {S}urveyor aerobraking period}.{\BBCQ}
\newblock
\APACjournalVolNumPages{Icarus}{192}{2}{327-347}.
\newblock
\begin{APACrefDOI} \doi{10.1016/j.icarus.2007.08.009} \end{APACrefDOI}
\PrintBackRefs{\CurrentBib}

\bibitem [\protect \citeauthoryear {%
Mart{\'i}nez%
\ \protect \BOthers {.}}{%
Mart{\'i}nez%
\ \protect \BOthers {.}}{%
{\protect \APACyear {2017}}%
}]{%
martinez17}
\APACinsertmetastar {%
martinez17}%
\begin{APACrefauthors}%
Mart{\'i}nez, G\BPBI M.%
, Newman, C\BPBI N.%
, De~Vicente-Retortillo, A.%
, Fischer, E.%
, Renno, N\BPBI O.%
, Richardson, M\BPBI I.%
\BDBL {}Vasavada, A\BPBI R.%
\end{APACrefauthors}%
\unskip\
\newblock
\APACrefYearMonthDay{2017}{Oct}{01}.
\newblock
{\BBOQ}\APACrefatitle {The Modern Near-Surface Martian Climate: A Review of
  In-situ Meteorological Data from {Viking to Curiosity}} {The modern
  near-surface martian climate: A review of in-situ meteorological data from
  {Viking to Curiosity}}.{\BBCQ}
\newblock
\APACjournalVolNumPages{Space Science Reviews}{212}{1}{295--338}.
\newblock
\begin{APACrefDOI} \doi{10.1007/s11214-017-0360-x} \end{APACrefDOI}
\PrintBackRefs{\CurrentBib}

\bibitem [\protect \citeauthoryear {%
Montabone%
\ \protect \BOthers {.}}{%
Montabone%
\ \protect \BOthers {.}}{%
{\protect \APACyear {2015}}%
}]{%
montabone15}
\APACinsertmetastar {%
montabone15}%
\begin{APACrefauthors}%
Montabone, L.%
, Forget, F.%
, Millour, E.%
, Wilson, R\BPBI J.%
, Lewis, S\BPBI R.%
, Cantor, B.%
\BDBL {}Wolff, M\BPBI J.%
\end{APACrefauthors}%
\unskip\
\newblock
\APACrefYearMonthDay{2015}{}{}.
\newblock
{\BBOQ}\APACrefatitle {Eight-year climatology of dust optical depth on {M}ars}
  {Eight-year climatology of dust optical depth on {M}ars}.{\BBCQ}
\newblock
\APACjournalVolNumPages{Icarus}{251}{}{65-95}.
\newblock
\begin{APACrefDOI} \doi{10.1016/j.icarus.2014.12.034} \end{APACrefDOI}
\PrintBackRefs{\CurrentBib}

\bibitem [\protect \citeauthoryear {%
Montabone%
, Lewis%
, Read%
\BCBL {}\ \BBA {} Hinson%
}{%
Montabone%
\ \protect \BOthers {.}}{%
{\protect \APACyear {2006}}%
}]{%
montabone06}
\APACinsertmetastar {%
montabone06}%
\begin{APACrefauthors}%
Montabone, L.%
, Lewis, S\BPBI R.%
, Read, P\BPBI L.%
\BCBL {}\ \BBA {} Hinson, D\BPBI P.%
\end{APACrefauthors}%
\unskip\
\newblock
\APACrefYearMonthDay{2006}{}{}.
\newblock
{\BBOQ}\APACrefatitle {Validation of {M}artian meteorological data assimilation
  for {MGS/TES} using radio occultation measurements} {Validation of {M}artian
  meteorological data assimilation for {MGS/TES} using radio occultation
  measurements}.{\BBCQ}
\newblock
\APACjournalVolNumPages{Icarus}{185}{1}{113-132}.
\newblock
\begin{APACrefDOI} \doi{10.1016/j.icarus.2006.07.012} \end{APACrefDOI}
\PrintBackRefs{\CurrentBib}

\bibitem [\protect \citeauthoryear {%
Montabone%
\ \protect \BOthers {.}}{%
Montabone%
\ \protect \BOthers {.}}{%
{\protect \APACyear {2014}}%
}]{%
montabone14}
\APACinsertmetastar {%
montabone14}%
\begin{APACrefauthors}%
Montabone, L.%
, Marsh, K.%
, Lewis, S\BPBI R.%
, Read, P\BPBI L.%
, Smith, M\BPBI D.%
, Holmes, J.%
\BDBL {}Pamment, A.%
\end{APACrefauthors}%
\unskip\
\newblock
\APACrefYearMonthDay{2014}{}{}.
\newblock
{\BBOQ}\APACrefatitle {The {Mars Analysis Correction Data Assimilation (MACDA)}
  Dataset V1.0} {The {Mars Analysis Correction Data Assimilation (MACDA)}
  dataset v1.0}.{\BBCQ}
\newblock
\APACjournalVolNumPages{Geosci. Data J.}{1}{2}{129-139}.
\newblock
\begin{APACrefDOI} \doi{10.1002/gdj3.13} \end{APACrefDOI}
\PrintBackRefs{\CurrentBib}

\bibitem [\protect \citeauthoryear {%
Mooring%
\ \BBA {} Wilson%
}{%
Mooring%
\ \BBA {} Wilson%
}{%
{\protect \APACyear {2015}}%
}]{%
mooring15}
\APACinsertmetastar {%
mooring15}%
\begin{APACrefauthors}%
Mooring, T\BPBI A.%
\BCBT {}\ \BBA {} Wilson, R\BPBI J.%
\end{APACrefauthors}%
\unskip\
\newblock
\APACrefYearMonthDay{2015}{}{}.
\newblock
{\BBOQ}\APACrefatitle {Transient eddies in the {MACDA} {M}ars reanalysis}
  {Transient eddies in the {MACDA} {M}ars reanalysis}.{\BBCQ}
\newblock
\APACjournalVolNumPages{J. Geophys. Res.}{120}{}{1671-1696}.
\newblock
\begin{APACrefDOI} \doi{10.1002/2015JE004824} \end{APACrefDOI}
\PrintBackRefs{\CurrentBib}

\bibitem [\protect \citeauthoryear {%
Mulholland%
, Lewis%
, Read%
, Madeleine%
\BCBL {}\ \BBA {} Forget%
}{%
Mulholland%
\ \protect \BOthers {.}}{%
{\protect \APACyear {2016}}%
}]{%
mulholland15}
\APACinsertmetastar {%
mulholland15}%
\begin{APACrefauthors}%
Mulholland, D\BPBI P.%
, Lewis, S\BPBI R.%
, Read, P\BPBI L.%
, Madeleine, J\BHBI B.%
\BCBL {}\ \BBA {} Forget, F.%
\end{APACrefauthors}%
\unskip\
\newblock
\APACrefYearMonthDay{2016}{}{}.
\newblock
{\BBOQ}\APACrefatitle {The solsticial pause on {M}ars: 2 Modelling and
  investigation of causes} {The solsticial pause on {M}ars: 2 modelling and
  investigation of causes}.{\BBCQ}
\newblock
\APACjournalVolNumPages{Icarus}{264}{}{465-477}.
\newblock
\begin{APACrefDOI} \doi{10.1016/j.icarus.2015.08.038} \end{APACrefDOI}
\PrintBackRefs{\CurrentBib}

\bibitem [\protect \citeauthoryear {%
Navarro%
\ \protect \BOthers {.}}{%
Navarro%
\ \protect \BOthers {.}}{%
{\protect \APACyear {2017}}%
}]{%
navarro17}
\APACinsertmetastar {%
navarro17}%
\begin{APACrefauthors}%
Navarro, T.%
, Forget, F.%
, Millour, E.%
, Greybush, S\BPBI J.%
, Kalnay, E.%
\BCBL {}\ \BBA {} Miyoshi, T.%
\end{APACrefauthors}%
\unskip\
\newblock
\APACrefYearMonthDay{2017}{}{}.
\newblock
{\BBOQ}\APACrefatitle {The Challenge of Atmospheric Data Assimilation on
  {M}ars} {The challenge of atmospheric data assimilation on {M}ars}.{\BBCQ}
\newblock
\APACjournalVolNumPages{Earth and Space Science}{4}{12}{690-722}.
\newblock
\begin{APACrefDOI} \doi{10.1002/2017EA000274} \end{APACrefDOI}
\PrintBackRefs{\CurrentBib}

\bibitem [\protect \citeauthoryear {%
Steele%
\ \protect \BOthers {.}}{%
Steele%
\ \protect \BOthers {.}}{%
{\protect \APACyear {2014}}%
}]{%
steele14}
\APACinsertmetastar {%
steele14}%
\begin{APACrefauthors}%
Steele, L\BPBI J.%
, Lewis, S\BPBI R.%
, Patel, M\BPBI R.%
, Montmessin, F.%
, Forget, F.%
\BCBL {}\ \BBA {} Smith, M\BPBI D.%
\end{APACrefauthors}%
\unskip\
\newblock
\APACrefYearMonthDay{2014}{}{}.
\newblock
{\BBOQ}\APACrefatitle {The seasonal cycle of water vapour on {M}ars from
  assimilation of {Thermal Emission Spectrometer} data} {The seasonal cycle of
  water vapour on {M}ars from assimilation of {Thermal Emission Spectrometer}
  data}.{\BBCQ}
\newblock
\APACjournalVolNumPages{Icarus}{237}{}{97 - 115}.
\newblock
\begin{APACrefDOI} \doi{10.1016/j.icarus.2014.04.017} \end{APACrefDOI}
\PrintBackRefs{\CurrentBib}

\bibitem [\protect \citeauthoryear {%
Wang%
\ \BBA {} Ingersoll%
}{%
Wang%
\ \BBA {} Ingersoll%
}{%
{\protect \APACyear {2003}}%
}]{%
wang03c}
\APACinsertmetastar {%
wang03c}%
\begin{APACrefauthors}%
Wang, H.%
\BCBT {}\ \BBA {} Ingersoll, A\BPBI P.%
\end{APACrefauthors}%
\unskip\
\newblock
\APACrefYearMonthDay{2003}{}{}.
\newblock
{\BBOQ}\APACrefatitle {Cloud-tracked winds for the first {Mars Global Surveyor}
  mapping year} {Cloud-tracked winds for the first {Mars Global Surveyor}
  mapping year}.{\BBCQ}
\newblock
\APACjournalVolNumPages{Journal of Geophysical Research: Planets}{108}{E9}{}.
\newblock
\begin{APACrefDOI} \doi{10.1029/2003JE002107} \end{APACrefDOI}
\PrintBackRefs{\CurrentBib}

\bibitem [\protect \citeauthoryear {%
Waugh%
\ \protect \BOthers {.}}{%
Waugh%
\ \protect \BOthers {.}}{%
{\protect \APACyear {2016}}%
}]{%
waugh16}
\APACinsertmetastar {%
waugh16}%
\begin{APACrefauthors}%
Waugh, D\BPBI W.%
, Toigo, A\BPBI D.%
, Guzewich, S\BPBI D.%
, Greybush, S\BPBI J.%
, Wilson, R\BPBI J.%
\BCBL {}\ \BBA {} Montabone, L.%
\end{APACrefauthors}%
\unskip\
\newblock
\APACrefYearMonthDay{2016}{}{}.
\newblock
{\BBOQ}\APACrefatitle {Martian polar vortices: Comparison of reanalyses}
  {Martian polar vortices: Comparison of reanalyses}.{\BBCQ}
\newblock
\APACjournalVolNumPages{Journal of Geophysical Research:
  Planets}{121}{9}{1770-1785}.
\newblock
\begin{APACrefDOI} \doi{10.1002/2016JE005093} \end{APACrefDOI}
\PrintBackRefs{\CurrentBib}

\bibitem [\protect \citeauthoryear {%
Wilks%
}{%
Wilks%
}{%
{\protect \APACyear {2019}}%
{\protect \APACexlab {{\protect \BCnt {1}}}}}]{%
wilks_ch4}
\APACinsertmetastar {%
wilks_ch4}%
\begin{APACrefauthors}%
Wilks, D\BPBI S.%
\end{APACrefauthors}%
\unskip\
\newblock
\APACrefYearMonthDay{2019{\protect \BCnt {1}}}{}{}.
\newblock
{\BBOQ}\APACrefatitle {Chapter 4 - Parametric Probability Distributions}
  {Chapter 4 - parametric probability distributions}.{\BBCQ}
\newblock
\BIn{} \APACrefbtitle {Statistical Methods in the Atmospheric Sciences}
  {Statistical methods in the atmospheric sciences}\ (\PrintOrdinal{Fourth}\
  \BEd, \BPG~77-141).
\newblock
\APACaddressPublisher{}{Elsevier}.
\newblock
\begin{APACrefDOI} \doi{10.1016/B978-0-12-815823-4.00004-3} \end{APACrefDOI}
\PrintBackRefs{\CurrentBib}

\bibitem [\protect \citeauthoryear {%
Wilks%
}{%
Wilks%
}{%
{\protect \APACyear {2019}}%
{\protect \APACexlab {{\protect \BCnt {2}}}}}]{%
wilks_ch5}
\APACinsertmetastar {%
wilks_ch5}%
\begin{APACrefauthors}%
Wilks, D\BPBI S.%
\end{APACrefauthors}%
\unskip\
\newblock
\APACrefYearMonthDay{2019{\protect \BCnt {2}}}{}{}.
\newblock
{\BBOQ}\APACrefatitle {Chapter 5 - Frequentist Statistical Inference} {Chapter
  5 - frequentist statistical inference}.{\BBCQ}
\newblock
\BIn{} \APACrefbtitle {Statistical Methods in the Atmospheric Sciences}
  {Statistical methods in the atmospheric sciences}\ (\PrintOrdinal{Fourth}\
  \BEd, \BPG~143-207).
\newblock
\APACaddressPublisher{}{Elsevier}.
\newblock
\begin{APACrefDOI} \doi{10.1016/B978-0-12-815823-4.00005-5} \end{APACrefDOI}
\PrintBackRefs{\CurrentBib}

\bibitem [\protect \citeauthoryear {%
Wilson%
, Lewis%
, Montabone%
\BCBL {}\ \BBA {} Smith%
}{%
Wilson%
\ \protect \BOthers {.}}{%
{\protect \APACyear {2008}}%
}]{%
wilson08}
\APACinsertmetastar {%
wilson08}%
\begin{APACrefauthors}%
Wilson, R\BPBI J.%
, Lewis, S\BPBI R.%
, Montabone, L.%
\BCBL {}\ \BBA {} Smith, M\BPBI D.%
\end{APACrefauthors}%
\unskip\
\newblock
\APACrefYearMonthDay{2008}{}{}.
\newblock
{\BBOQ}\APACrefatitle {Influence of water ice clouds on {M}artian tropical
  atmospheric temperatures} {Influence of water ice clouds on {M}artian
  tropical atmospheric temperatures}.{\BBCQ}
\newblock
\APACjournalVolNumPages{Geophys. Res. Lett.}{35}{7}{L07202}.
\newblock
\begin{APACrefDOI} \doi{10.1029/2007GL032405} \end{APACrefDOI}
\PrintBackRefs{\CurrentBib}

\bibitem [\protect \citeauthoryear {%
Zhao%
, Greybush%
, Wilson%
, Hoffman%
\BCBL {}\ \BBA {} Kalnay%
}{%
Zhao%
\ \protect \BOthers {.}}{%
{\protect \APACyear {2015}}%
}]{%
zhao15}
\APACinsertmetastar {%
zhao15}%
\begin{APACrefauthors}%
Zhao, Y.%
, Greybush, S\BPBI J.%
, Wilson, R\BPBI J.%
, Hoffman, R\BPBI N.%
\BCBL {}\ \BBA {} Kalnay, E.%
\end{APACrefauthors}%
\unskip\
\newblock
\APACrefYearMonthDay{2015}{}{}.
\newblock
{\BBOQ}\APACrefatitle {Impact of assimilation window length on diurnal features
  in a {M}ars atmospheric analysis} {Impact of assimilation window length on
  diurnal features in a {M}ars atmospheric analysis}.{\BBCQ}
\newblock
\APACjournalVolNumPages{Tellus A}{67}{}{26042}.
\newblock
\begin{APACrefDOI} \doi{10.3402/tellusa.v67.26042} \end{APACrefDOI}
\PrintBackRefs{\CurrentBib}

\end{thebibliography}

\end{document}